\documentclass[a4paper,fleqn,usenatbib]{mnras}

\usepackage{newtxtext,newtxmath}
\usepackage{hyperref}
\usepackage[T1]{fontenc}
\usepackage{ae,aecompl}

\usepackage{graphicx}	% Including figure files
\usepackage{amsmath}	% Advanced maths commands
\usepackage{amssymb}	% Extra maths symbols
\newcommand{\sigsfr}{$\Sigma_{\rm SFR}$}
\newcommand{\sigsfrU}{M$_{\odot}$ yr$^{-1}$ kpc$^{-2}$}
\newcommand{\arc}{^{\prime\prime}}
\newcommand{\kms}{\,km\,s$^{-1}$}
\newcommand{\paa}{Pa-$\alpha$}
\newcommand{\lpaa}{L(Pa-$\alpha$)}
\newcommand{\vs}{$V_{\rm shear}$}
\newcommand{\pan}{\textsc{pan}}
\newcommand{\idl}{\textsc{idl}}
\newcommand{\flux}{erg s$^{-1}$ cm$^{-2}$}

\newcommand{\gbk}{\textsc{gbkfit}}
\newcommand{\sigp}{$\sigma_{\rm peaks}$}

\title[Velocity dispersion of turbulent galaxies]{The connection between the peaks in velocity dispersion and star-forming clumps of turbulent galaxies}

\author[Oliva-Altamirano et al.]{
P.~{Oliva-Altamirano},$^{1}$\thanks{E-mail: poliva@astro.swin.edu.au} D.~{Fisher},$^{1}$ K.~{Glazebrook},$^{1}$ E.~{Wisnioski},$^{2}$ G.~{Bekiaris},$^{1}$ \newauthor{} R.~{Bassett},$^{1}$ D.~{Obreschkow},$^{3}$ and R.~{Abraham},$^{4}$
\\
$^{1}$Centre for Astrophysics \& Supercomputing, Swinburne University of Technology, Hawthorn, VIC 3122, Australia\\
$^{2}$Max-Planck-Institut für extraterrestrische Physik (MPE), Scheinerstrasse 1, D-85748 Garching, Germany\\
$^{3}$International Centre for Radio Astronomy Research, University of Western Australia, 7 Fairway, Crawley, WA 6009, Australia\\
$^{4}$Department of Astronomy and Astrophysics, University of Toronto, 50 St George Street, Toronto, ON M5S 3H8, Canada
}

\date{Accepted XXX. Received YYY; in original form ZZZ}

\pubyear{2016}

\begin{document}
\label{firstpage}
\pagerange{\pageref{firstpage}--\pageref{lastpage}}
\maketitle

\begin{abstract}
We present Keck/OSIRIS  adaptive optics observations with 150-400 pc spatial sampling of 7 turbulent, clumpy disc galaxies from the DYNAMO sample ($0.07<z<0.2$). DYNAMO galaxies have previously been shown to be well matched in properties to main sequence galaxies at $z\sim1.5$. Integral field spectroscopy observations using adaptive optics are subject to a number of systematics including a variable PSF and spatial sampling, which we account for in our analysis. We present gas velocity dispersion maps corrected for these effects, and confirm that DYNAMO galaxies do have high gas velocity dispersion ($\sigma=40-80$\kms), even at high spatial sampling.
We find statistically significant structure in 6 out of 7 galaxies. The most common distance between the peaks in velocity dispersion (\sigp) and emission line peaks is $\sim0.5$~kpc, we note this is very similar to the average size of a clump measured with HST H$\alpha$ maps. This could suggest that \sigp~in clumpy galaxies likely arise due to some interaction between the clump and the surrounding ISM of the galaxy, though our observations cannot distinguish between outflows, inflows or velocity shear. Observations covering a wider area of the galaxies will be needed to confirm this result. 
\end{abstract}


%%%%%%%%%%%%%%%%%%%%%%%%%%%%%%%%%%%%%%%%%%%%%%%%%%
%%%%%%%%%%%%%%%%% BODY OF PAPER %%%%%%%%%%%%%%%%%%
%==========================================================================================
\section{Introduction}
The most active epoch in star formation in the Universe occurs at $1<z<3$ \citep{madau2014}. Star forming galaxies during this epoch show very significant differences from $z\approx0$ spirals. For example $z>1$ galaxies have disturbed rest frame-UV morphologies \citep{Abraham_1996, Conselice_2000, Elmegreen_2005}, and extremely bright knots of star formation (so-called ``clumps") that are  100-1000$\times$ brighter than local star forming regions \citep{Genzel_2011,Swinbank_2012,Wisnioski_2011}.  Adaptive Optic (AO) observations revealed that these objects can be classified in three groups: mergers, dispersion-dominated galaxies and turbulent discs with elevated velocity dispersions ($V_{\rm shear}$/$\sigma_{\rm gal}\sim1-4$) and all three groups show high velocity dispersions \citep[$\sigma_{\rm gal} > 50$~\kms; e.g.][for a review]{Law_2009,Jones_2010, Genzel_2011, Wisnioski_2012, Swinbank_2012,Glazebrook_2013}.  Turbulent disc galaxies are distinct from local Universe discs. Local Universe disc galaxies have lower gas velocity dispersion ($\sigma_{\rm gal} \sim 25$~\kms) and low gas fractions ($f_{gas}<10$\%), conversely high-$z$ discs have higher gas velocity dispersions ($\sim 40-100$~\kms) and high molecular gas fractions ($f_{gas}\sim20-60$\%). 

Models of unstable disc formation suggest that turbulent motions in the interstellar medium would cause high velocity dispersion. \citep{Immeli_2004,Immeli_2004b,Bournard_2008,Elmegreen_2008}. Recently, \cite{Fisher2017} show that the size of clumps in turbulent discs linearly correlates disc kinematics, and agrees with predictions of unstable disc galaxy evolution. Similarly, \cite{white2017} show that gas fractions and disc kinematics also relate in a manner that is consistent with predictions of unstable (or marginally stable) disc galaxies. 

A detailed picture of the role of the turbulence in the formation and life-cycle of clumps is still unknown. Some simulations suggest that the turbulent motions caused by radiative feedback and gas accretion in galaxies can maintain the high velocity dispersion required to regulate the stability in discs \citep{Ceverino_2009,Dekel_2009}. Alternatively, \cite{krumholz2016} argue that galaxy averaged observations are most consistent with a gravity-only mechanism driving high velocity dispersions. 

High-redshift observations on the spatial variation of velocity dispersions within turbulent discs are limited by both resolution and signal-to-noise (S/N), and thus poorly constrain the simulations mentioned above. At $z\sim2$, $0.\arc12$ correspond to a physical resolution of $\sim1000$ pc. At this resolution, the beam-smearing effects are significant, which makes the velocity dispersion measurements of clumps difficult to interpret. These observations are also marked by decreased flux at the detector, dropping the S/N significantly in areas where there is no active star formation \citep{Law_2006}. This effect artificially increases velocity dispersion in regions of low flux. Structure in velocity dispersion maps has been discussed in some observations \citep{Genzel_2011, Wisnioski_2011, Livermore_2015}, however, the associated large uncertainties, discussed above, have hindered interpretation and detailed analysis.

The DYnamics of Newly-Assembled Massive Objects \citep[DYNAMO;][]{Green_2014}\footnote{http://dynamo.swinburne.edu.au/index.html} survey is a sample of rare galaxies (space density of $\sim10^{-7}$Mpc$^{-3}$) at $z\sim0.1$ that are well matched in properties to $z\sim1$ main sequence galaxies. The DYNAMO galaxies have high fractions of gas \citep{Fisher_2014,white2017}, high SFRs ($1-40$~M$_\odot$ yr$^{-1}$ ) and high H$\alpha$ velocity dispersions \citep[$\sigma>50$\kms;][]{Green_2010, Green_2014,Bassett_2014}. They are morphologically similar to high redshift galaxies and pass the commonly used quantified definition of a ``clumpy galaxy" \citep{Fisher_2016}. Under the same observing conditions using AO enabled observations of emission lines one can achieve a resolution $5-10$ times better than at high redshifts. These galaxies, therefore, provide an excellent opportunity to study the well-resolved kinematics of turbulent discs.  

In this work we will use DYNAMO survey galaxies to explore the existence of structure in velocity dispersion distributions of turbulent discs, to later study the connection between this structure and the star-forming regions in the galaxy. In Section 2 we present the DYNAMO survey and DYNAMO-OSIRIS sample, analysed here. Section 3 summarizes the analysis: clump identification, emission line maps and kinematic measurements. Section 4 presents our results which are discussed in Section 5. Our conclusions are presented in Section 6. We use the cosmology H$_0 = 70$~km~s$^{{-}1}$~Mpc$^{{-}1}$, $\Omega_{\rm M} = 0.3$, $\Omega_\Lambda = 0.7$.

\section{Data}
\subsection{Sample selection}

The DYNAMO sample \citep{Green_2014} is selected from the Sloan Digital Sky Survey Data Release 4 \citep[SDSS;][]{Adelman_2006}. The sample consists of star-forming galaxies with no Active Galactic Nuclei (AGN) activity. The redshift range is designed to avoid the overlap between the sky emission and the H$\alpha$ line, sampling the full luminosity range between $0.055<z<0.084$ and the upper luminosity range between $0.129<z<0.254$. The primary objective of DYNAMO is to study low redshift analogous of high redshift turbulent galaxies ($z\sim1.5$). \citet[][]{Green_2014} gives a full description of the sample and presents the H$\alpha$ fluxes and initial kinematics of the DYNAMO galaxies. They use observations from the SPIRAL \citep{Sharp_2006}  and WiFES \citep{Dopita_2007} Integral Field Spectrograph (IFS) on the Anglo-Australian Telescope (AAT) and the ANU 2.3m Telescope, respectively. They present an initial kinematic classification of the objects. However, the low spatial sampling of these IFS ($\sim 0.\arc7 - 1.\arc0$ per spaxel) limits the possibility of studying the kinematics of the substructure within the galaxies.

To follow up on the results of \citet[][]{Green_2014}, we observed a  subset of 21 DYNAMO galaxies  with higher spectral and spatial sampling. This sub sample was observed with OSIRIS IFS \citep[OH Suppressing Infrared Imaging Spectrograph;][]{Larkin_2006} on Keck. Henceforth, we refer to it as DYNAMO-OSIRIS to avoid confusion with the parent sample.  

The main limitation of the sample selection was the adaptive optics (AO) requirement. i.e. the availability of at least one telluric star nearby the target galaxy. Ten of these objects are described in \citep{Green_2014} and eleven were subsequently added to the DYNAMO sample due to their proximity to adaptive optics laser guide stars. These new targets met all early selection criteria for DYNAMO targets, yet had not been included in \cite{Green_2014}. 

DYNAMO-OSIRIS includes objects that were classified as turbulent discs with high velocity dispersions ($\sigma \geq 40$\kms), blue colours ($g - i \leq 1$), M$_{star}\sim 5\times10^{10}$~M$_{\odot}$, and high SFRs ($10-40$ M$_\odot$ yr$^{-1}$). We aim to test the result found at high redshifts where large disc dispersions leads to large star-forming regions \citep{Genzel_2011, Wisnioski_2011}. The blue colours and high SFR preferentially select galaxies with dominant young active star formation. 

Three of the DYNAMO-OSIRIS galaxies overlap with the DYNAMO-HST \citep{Fisher_2016} subsample which consist of ten galaxies with H$\alpha$ imaging from the Hubble Space Telescope (HST), and eight overlap with the DYNAMO-Gemini subsample \citep{Bassett_2014,Obreschkow_2015} which consist of GMOS deep IFS observations of $\sim$fifteen objects. This overlap allows us to pursue high-resolution multi-wavelength studies of turbulent galaxies. From the 21 galaxies observed, only 7 satisfy the S/N$\geq4$ and disc-like kinematics required for this study.

The final sample of 7 galaxies presented here have SFR~$=5-42$~M$\odot$~yr$^{-1}$, and stellar masses $=1-6\times10^{10}$~M$\odot$ (\autoref{tab:obs}). As we will show in this work, their emission line morphology and kinematics are consistent with clumpy, turbulent discs. 3 out of the 7 galaxies: D13-5, G20-2, and G4-1 overlap with DYNAMO-HST \citep{Fisher_2016} and 4 out of the 7 galaxies: D13-5, G20-2, G4-1 and SDSS013527-1039 overlap with DYNAMO-GEMINI \citep{Bassett_2014,Obreschkow_2015}. 

\subsection{OSIRIS Observations}
We observe the DYNAMO-OSIRIS targets with AO enabled observations of the \paa~emission line with Keck/OSIRIS \citep[][]{Larkin_2006}. OSIRIS is an IFS organized in lenslet arrays with a $2048\times2048$ Hawaii-2 detector. It has a spectral resolution of R~$\sim3000$ and a spatial sampling of $0.\arc1$. The observations were taken in 2012 (July, September) with the Keck~I AO system \citep{Wizinowich_2006} in laser guide star mode (LGSAO). The seeing range was $0.\arc35 - 0.\arc80$.

We first acquired tip-tilt stars within the LGSAO field-of-view (FOV). We use 60 sec integrations of these tip-tilt stars to calculate the PSF of the observations. For the science observations, we take 4 exposure sequences of 900 sec, dithered by   $0.\arc05$ around the base point, to avoid bad pixels and cosmic ray contamination. For the sky frames, we need to offset the telescope from the object, as the galaxies commonly covered the whole FOV. Each sky frame was exposed for 30 minutes. The 7 galaxies in our sample were observed with the filters Kn1, Kn2 and Kc3, wavelength ranges $1960-2040$, $2040-2120$ and $2120-2220$~nm, respectively. These short ranges cover solely the \paa~line emission. The spectroscopic observations of the DYNAMO-OSIRIS galaxies are summarized in \autoref{tab:obs}.

\subsection{Data reduction and flux calibration}
\begin{table*}
	\centering
	\caption{Summary of OSIRIS observations. Column 1 shows the DYNAMO or SDSS ID of the galaxy. The object exposure sequence is 4$\times$900 sec and the sky exposure sequence is 2$\times$900 sec. The observations were taken in laser guide star mode. Column 2 shows the galaxy stellar mass \citep{Kauffmann_2003}. Column 3 shows the SFR from H$\alpha$ SPIRAL/WiFES observations \citep{Green_2014}. Column 9 shows the physical resolution corresponding to $0.\arcsec1$. Column 10 presents the overlap with other DYNAMO samples, H: HST, G: Gemini.}
	\label{tab:obs}
	\begin{tabular}{lccccccccccc}
		\hline
		ID & M$_*$& SFR$_{H\alpha}$ & RA & Dec & $z$ & Obs.&Filter&Resolution&Overlap\\
		             &[$10^{10}$ M$\odot$]&[M$_{\odot}$ yr$^{-1}$]&[J2000]&[J2000]&&Date&&[pc]&\\
		\hline
		  D13-5& 5.3& 21.20&13:30:07.009&00:31:53.450&0.0754&2012/07&Kn1&150&H,G\\
		  G20-2&2.2&17.27&20:44:02.922&-06:46:57.940&0.1411&2012/07&Kc3&260&H,G\\
		  G4-1&6.5&41.61&04:12:19.710&-05:54:48.670&0.1298&2012/09&Kn2&240&H,G\\
		  E23-1&2.0&15.99&23:58:33.897&14:54:45.358&0.1449&2012/07&Kc3&260&---\\
		  SDSS152159+0640&1.1&11.30&15:21:59.117&06:40:50.330&0.1826&2012/07&Kc3&320&---\\
		  SDSS212912-0734&1.1&4.95&21:29:12.151&-07:34:57.669&0.1835&2012/09&Kc3&320&---\\
		  SDSS013527-1039&1.1&5.95&01:35:27.100&-10:39:38.519&0.1269&2012/09&Kn2&230&G\\
		\hline
	\end{tabular}
\end{table*}
We use the OSIRIS data reduction pipeline version 2.3. The pipeline removes crosstalk, detector glitches and cosmic rays per frame, to later combine the data into a cube (two spatial coordinates $x$ and $y$, and one spectral coordinate $z$). The sky subtraction is achieved in two steps, first by a spatial nodding on the sky and later by custom \idl~routines based on the method of \citet{Davies_2007}.

To flux-calibrate the OSIRIS spectra, we use the telluric stars observed each night (an average of 3 telluric stars per night). We use a synthetic spectrum of the star to correct the shape of the observed spectrum and the density flux obtained from the 2MASS catalogue \citep{Skrutskie_2006} to calculate a scaling factor. With these two parameters, we compute a correction curve. The observed spectrum is flux calibrated by applying the correction curve. This is particularly easy in the near-Infrared as the stars are smooth in this region of the spectrum.

We compute a correction curve for each night of the observations and correct each 1D spectra in the galaxy's data cube. We check our results by integrating the corrected flux of each DYNAMO-OSIRIS galaxy and comparing the total magnitude with the Ks magnitude in the 2MASS catalogue. Our method corresponds to a first order flux calibration consistent with the 2MASS magnitudes within 20-30\%. This could be due to aperture effects since the OSIRIS FOV is in some cases smaller than the galaxies and might be missing stellar light. This accuracy is acceptable for the purpose of our study. The flux calibration is internally consistent for each map, and therefore sufficient for studying substructure.

\section{Analysis}
The aim of this work is the investigation of kinematic substructure in turbulent discs at $150-400$~pc sampling. To do this we compare the velocity dispersion in the star-forming regions (clumps) with the velocity dispersion of the kinematic disc. We use the \paa~emission line to measure the gas kinematics as well as the SFR across each galaxy. 

\subsection{Emission-line maps and star formation rates} \label{sec:sfr}
The star-forming regions, in general, have high S/N (\paa~peak S/N $> 30$). However, we need to ensure a good S/N in regions of low emission as well. To improve the general S/N, we apply a median filter to the galaxy cube, i.e. we replace the value of a given spaxel with the median of the adjacent spaxels. Later, we remove the spaxels with \paa~peak S/N $<4$. This enables a clean measurement of the inner galaxy properties.

We calculate the \paa~flux, per spaxel, per galaxy by fitting a gaussian to the line emission. From the \paa~fluxes we calculate the \lpaa~and use the following equations to compute the SFRs:
\begin{equation}
L(\text{Pa-}\alpha)_{\text{corr}} =  10^{0.4A_{\text{Pa-}\alpha}}L(\text{Pa-}\alpha) 
\end{equation}
\begin{equation}
A_{\text{Pa-}\alpha}=0.52 ~log \left( \frac{\text{Pa-}\alpha/\text{H}\alpha}{0.128}\right) 
\label{eq:dust}
\end{equation}
\begin{equation}
\text{SFR} =  5.53\times 10^{-42} \times0.128\times L(\text{Pa-}\alpha)_{\text{corr}}
\label{eq:sfr}
\end{equation}
where $A_{\rm{Pa-}\alpha}$ (\autoref{eq:dust}) is the dust correction factor from \citet{Calzetti_2000}, $\text{Pa-}\alpha/\text{H}\alpha$ (\autoref{eq:sfr}) is the flux \paa-to- H$\alpha$ ratio, and the SFR is the relation for H$\alpha$ from \citet{Hao_2011} scaled by the intrinsic \paa-to- H$\alpha$ ratio from Case B recombination, 0.128. 

The effect of dust in the near-Infrared is almost negligible which makes $A_{\rm{Pa-}\alpha}$ small for our observations, mean $A_{\rm{Pa-}\alpha}=0.12\pm0.01$. $A_{\rm{Pa-}\alpha}$ was calculated from the \paa/H$\alpha$ ratio by: a) comparing the \paa~OSIRIS observations with the H$\alpha$ SPIRAL/WiFES observations of \citet{Green_2014}; b) comparing with H$\alpha$ HST observations \citep{Bassett_2016}. Both methods revealed similar results. 

The \paa~fluxes could be affected by AO PSF effects. The AO system distributes the light from a point source in large wings, larger than the PSF core. \citet{Bassett_2016} found, by comparing OSIRIS and HST emission line maps, that D13-5, G20-2 and G4-1 \paa~fluxes need corrections in the range of 10-30\% for fluxes with radii smaller than 0.3 arcsec, for structures larger than this there is negligible difference.  This level of uncertainty is acceptable for the purpose of our analysis, in which the emission line map is mainly used to identify clump locations. We present our \paa~emission-line maps in units of \sigsfr~(\sigsfrU), to compare across galaxies without being biased by the different physical scales. 

\subsection{Clump identification}
\begin{figure}
     \includegraphics[width=\columnwidth]{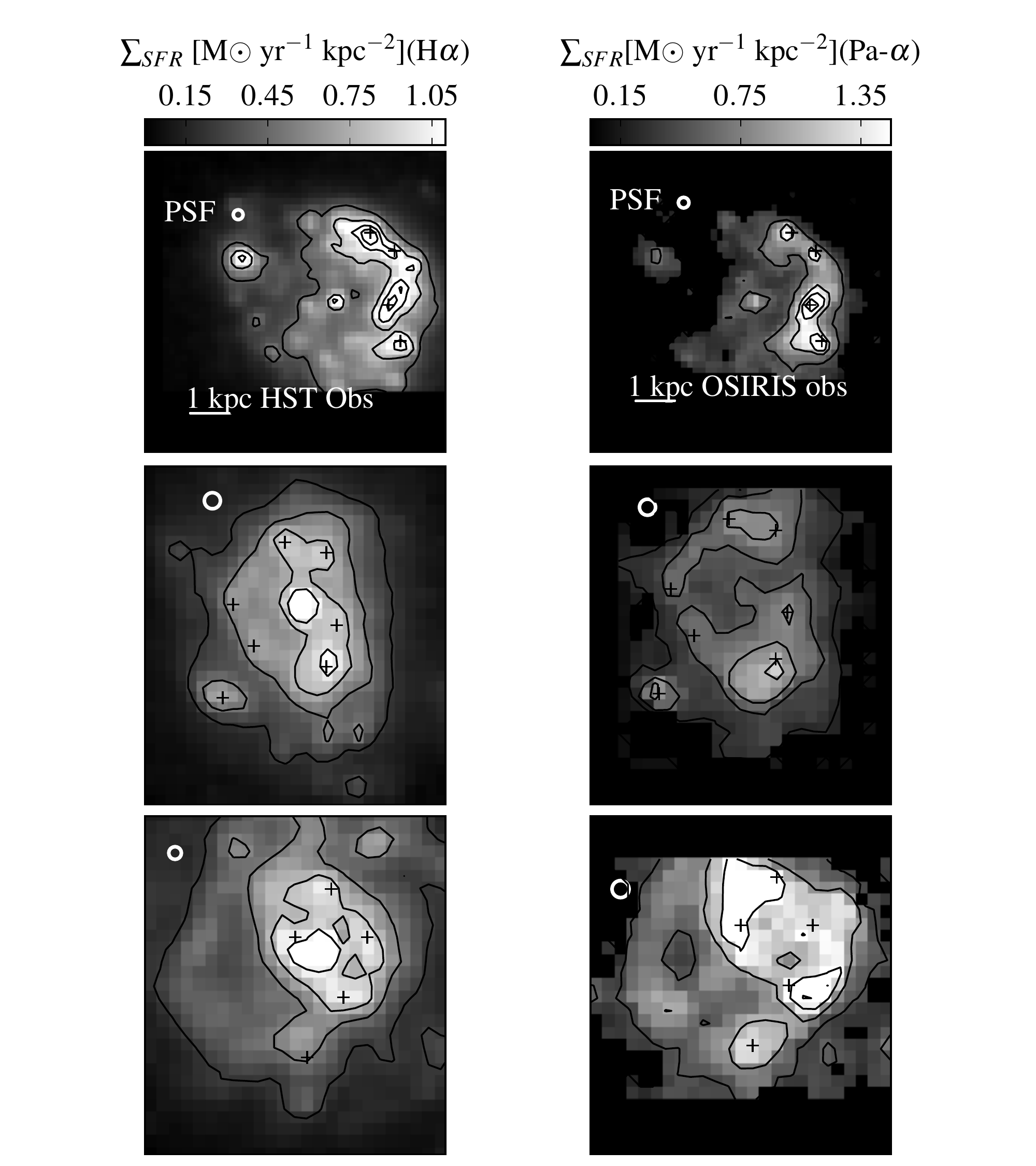}
    \caption{Comparison between the H$\alpha$ and the \paa~map, galaxies D13-5 (upper panels), G20-2 (middle panels), and G4-1 (lower panels). The beam size is shown as a white circumference in each panel. The left-hand panels are the HST H$\alpha$ maps degraded to OSIRIS spatial sampling ($0.\arc1$). The right-hand panels show the \paa~maps from the OSIRIS observations. The \sigsfr~contours represent 20\% of the total \sigsfr. The crosses represent the identified clumps (2$\times$ brighter than the galaxy smoothed floor for \paa~and 3$\times$ brighter than the galaxy smoothed floor for H$\alpha$). The images are shown at the same physical scale to facilitate visual inspection. The clump positions are consistent between the H$\alpha$ and \paa~emission.
	}
    \label{fig:clumpid}
\end{figure}

Three of the galaxies in our sample (D13-5, G20-2, G4-1) also have HST observations of H$\alpha$. These galaxies were explored by \citet{Fisher_2016} to find an automatic method to identify clumps and measure their sizes. In \autoref{fig:clumpid}, we show the comparison between two HST H$\alpha$ maps degraded to the OSIRIS spatial sampling ($0.\arc1$) and the \paa~maps from the OSIRIS observations\footnote{For information about the registration of the images and a more detail comparison between the OSIRIS \paa~map and the HST H$\alpha$ map refer to \citet{Bassett_2016}.}. We find a one-to-one correspondence between the two emission lines. The clumps seen in the H$\alpha$ map are also visible in the \paa~map. 

To find the clump position, we follow the method implemented  by \citet{Fisher_2016}. This process recovers similar objects to those found in H$\alpha$ maps of high redshift galaxies \citep{Jones_2010, Genzel_2011, Livermore_2012, Guo_2015}. It consists of employing a simple unsharp masking technique. We first convolve the \paa~map with a Gaussian of $\sim1\arc$. Then we divide the full resolution image by the smoothed image. The clumps are the emission peaks that are 2$\times$ brighter than the smoothed map of the galaxy \citep[3$\times$ brighter when analysing H$\alpha$;][]{Fisher_2016}. Our sanity check is that we find in the \paa~map the same clumps found in the HST H$\alpha$ map by \citet{Fisher_2016}. 

The number of clumps in each galaxy ranges from 3 to 7 within the masked OSIRIS FOV (see \autoref{tab:gal_prop}). The total coverage of the galaxy varies from $2-4\arc$~(5-8 kpc) as we are limited by a S/N cut of 4 that will allow us to measure reliable kinematics. For most galaxies the mask region does not cover the whole galaxy.

\subsection{Gas kinematics}\label{sec:kin}
\begin{figure*}
 \includegraphics[width=0.8\linewidth]{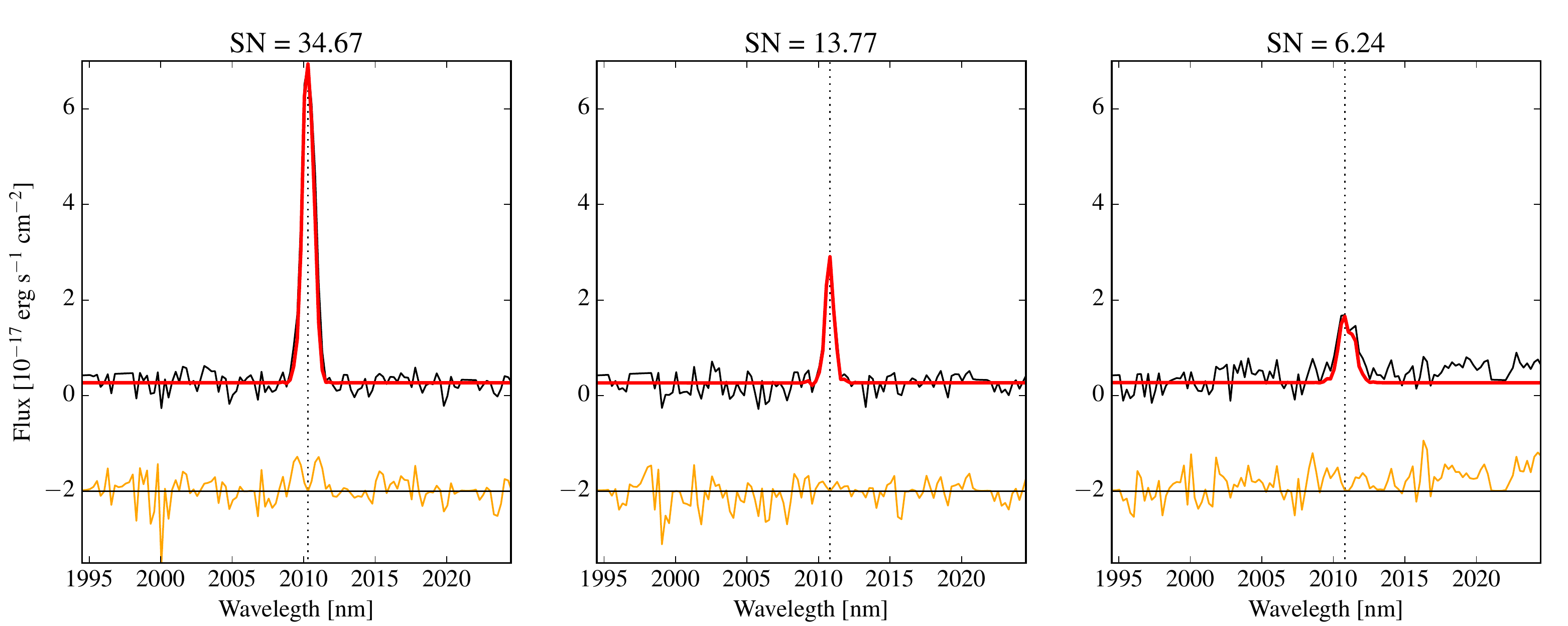}
 \caption{Gaussian fit (from \pan) of the observed spectra of galaxy D13-5. We present three typical peak S/N found in the OSIRIS observations. The black line represents the observed spectra, the red line represents the Gaussian fit form \pan~and the orange line represents the residuals of the fit as described in \autoref{sec:kin}. The \textbf{left-hand panel} shows a typical spectrum from the star-forming clumps. The \textbf{righ-hand panel} shows a typical spectrum found at the edges of the galaxies. We find a good fit in spaxels with S/N~$\geq4$. A colour version of this plot is available at the electronic edition.  
	}
\label{fig:fit}
\end{figure*}
We measure the \paa~line centroid and FWHM using the \idl~$\chi^2$ routine \pan~\citep[Peak ANalysis;][]{Dimeo_2005} modified by \citet{Westmoquette_2007} for astronomical purposes. \citet[][]{Westmoquette_2008,Westmoquette_2012} and \citet{Gruyters_2012} show that \pan~measures line-widths robustly, using a Levenberg-Marquardt technique to solve the least-square minimization problem. 

We input to \pan~the observed spectra and their associated error arrays (variance from the observations) provided by the OSIRIS data reduction pipeline. We fit a background polynomial and a Gaussian profile, to the \paa~emission line. To calculate the uncertainties we use the Monte-Carlo fitting provided by \pan. This routine repeats the fit 100 times by creating a synthetic spectrum from the best fit of the observed spectrum and the addition of Gaussian random errors. 

 \pan~provides three valuable checks to discriminate between good and bad fits, which we use in our analysis: a) The residuals of the fit, calculated as $r_i = (y_{i, \rm fit} - y_{i,\rm data})/\sigma_{i,\rm err}$. Where y$_{i,\rm data}$ is the observed spectrum; y$_{i, \rm fit}$ is the fit to the observed spectrum; $\sigma_{i,\rm err}$ is the associated error. From a good fit we expect a residual that resembles white noise. b) The values of the $\chi^2$. c) The uncertainties of the fit. We reject fits with uncertainties greater than the FWHM measured. In \autoref{fig:fit} we show the spectra of typical spaxels in galaxy D13-5 at three different peak S/N. The S/N~$> 30$ (left-hand panel) represents the star-forming regions (clumps). The S/N~$<6$ (right-hand panel) is commonly found at the edges of the galaxy where the light profile becomes faint. We find that the Gaussian fit is good in spaxels with S/N~$\geq4$.     

After fitting the \paa~emission line, we proceed to calculate the velocities from the centroid and the velocity dispersions from the FWHM of the Gaussian profile. 

We subtract in quadrature from the velocity dispersion the OSIRIS instrumental broadening, measured from the OH lines \citep{Rousselot_2000}. The OSIRIS instrumental broadening shows a standard variation of $\pm 6.4$\kms across the FOV.  We find that compared to our measured velocity dispersions this variation is small and does not affect our conclusions. We discuss this effect on our identification of velocity dispersion peaks in section 4.1. Thus, we use a constant instrumental resolution of $41$\kms~\citep[similar to ][]{Mieda_2016} across the FOV. This, however, does increase the uncertainty of our velocity dispersion measurements. We therefore add in quadrature 6.4\kms~to the measurement errors.

\subsubsection{AO PSF and typical noise effects on velocity dispersion measurements}
\begin{figure*}
\begin{center}
%\hspace*{1.9cm}
\includegraphics[width=0.8\linewidth]{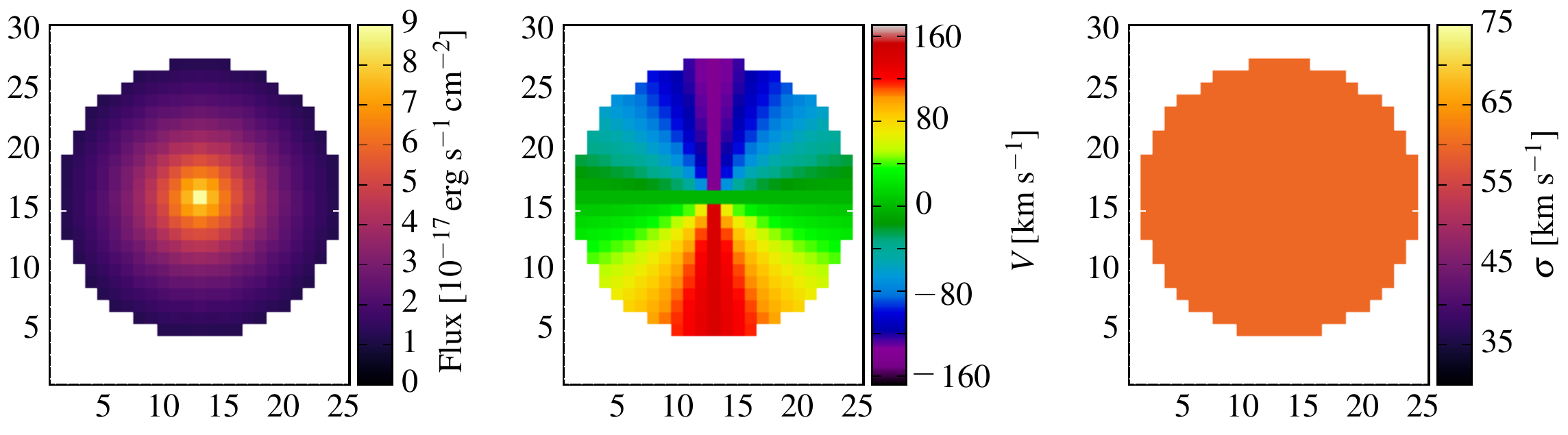}
%\hspace*{1.9cm}
\includegraphics[width=0.8\linewidth]{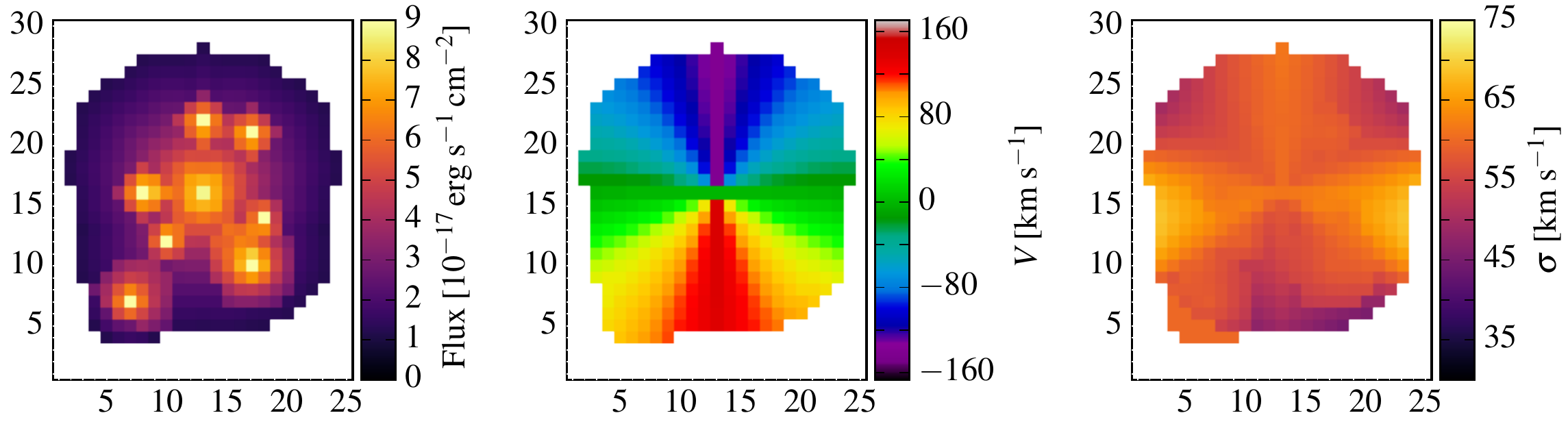}
%\hspace*{1.9cm}
\includegraphics[width=0.8\linewidth]{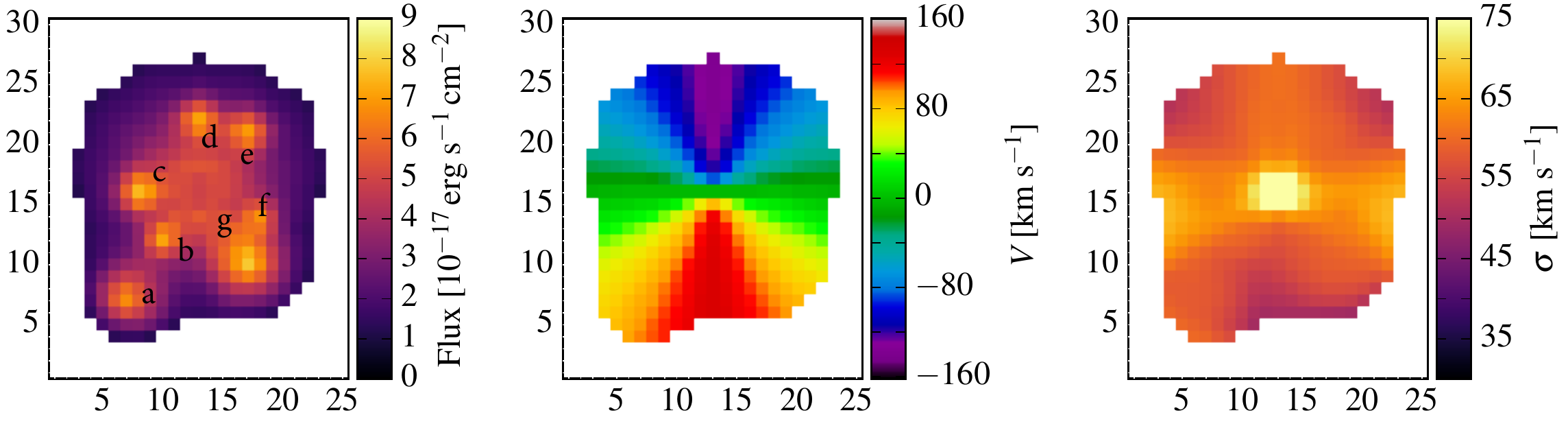}

\caption{AO PSF and noise simulations. The \textbf{upper panels} show the initial input of our simulations: a simple disc with exponential light profile (left panel), velocity map (middle panel) and the constant velocity dispersion map (60 \kms; right panel). The \textbf{middle panels} present the second stage of the simulation when we add 7 typical non-central clumps (reproducing the galaxy G20-2) to the exponential disc and the typical OSIRIS noise. Left panel shows the flux map, the middle and right panels show the kinematic maps recovered with \pan. The \textbf{bottom panels} show the flux map after convolving the galaxy with the widest AO PSF (Gaussian core $FWHM=3\pm0.51$ spaxels plus wings), and the recovered velocity and velocity dispersion maps from \pan. In this final stage we see the spatial sampling effect from the OSIRIS AO PSF at the centre and the noise effect as cross-like patterns in the velocity dispersion map. A colour version of this plot is available at the electronic edition. } %

\label{fig:sim}
\end{center}
\end{figure*}

As mentioned in \autoref{sec:sfr} the AO system creates a PSF that can not be modeled by a single Gaussian profile. The PSF profile of AO systems combine a central core and broader wings. These wings distribute the flux from a point source in a wider area and can also catch fluxes from surrounding sources. \citet{Bassett_2016} found that in our sample the PSF wings elevate the real flux in a given region of a galaxy up to $\sim50$\%. Here we seek to quantify the impact of a non-Gaussian profile PSF on velocity dispersion. We refer to this phenomenon as \textit{AO PSF effect} throughout the paper.

A second source of possible uncertainty in measuring velocity dispersions is noise. Elevated noise increases the recovered velocity dispersion, i.e. the Gaussian profile that describes the emission line would be distorted by  the noise patterns. This would widen the FWHM and artificially increase the velocity dispersion. This is a major problem for the kinematic studies of high redshift galaxies \citep[e.g.][]{Law_2009,Genzel_2011,Livermore_2015}.  

We use simulations to quantify the impact of the AO PSF effect and typical noise over our velocity dispersion maps. These simulations are divided into three stages that are described in the following paragraphs.

1) We create an empty box of the average size of the galaxies in our sample, $2.\arc5\times3\arc$. As first input, we add a simulated face-on galaxy with an exponential light profile. This is done by creating a simulated spectrum (without noise) per spaxel, containing a single emission line, \paa. We give the same velocity dispersion to all the spaxels and add in quadrature the OSIRIS instrumental broadening profile (shuffled randomly between 35 and 47\kms). The constant velocity dispersion map, facilitates the identification of any changes caused by the systematics.

We find that the instrumental resolution affects areas of raw velocity dispersion below 52\kms, which is equivalent to corrected velocity dispersions $<22.24$\kms. The majority of the spaxels in our galaxies have recovered velocity dispersions above 20\kms. Therefore, the OSIRIS instrumental resolution does not affect our conclusions.

Finally, we apply a disc rotation model to the galaxy: a simple \textit{arctan} profile \citep{Courteau_1997}. The upper panels of \autoref{fig:sim} show the exponential light profile (left panel), the initial velocity map (middle panel) and an initial velocity dispersion of $\sigma_{\rm input}=60$~\kms (left panel).

2) The second step is to create a simulated galaxy similar to those in our sample, to do so we add star-forming clumps to the disc in randomized positions. These clumps are symmetric, exponential light profiles \footnote{To reproduce the physical shapes of clumps (e.g. elongated, asymmetrical) would be beyond the scope of the simulation.}. The clump sizes vary from $0.\arc2 -0.\arc6$ (2~-~6 spaxels) and fluxes span from 1~-~5\% of the total flux of the galaxy \citep[][]{Wuyts_2012,Guo_2015, Fisher_2016}. 

Later we add to the simulated spectra random noise of the order of the OSIRIS observations. To calculate the typical noise we chose three spaxels per galaxy from a region dominated by the sky, where galaxy flux is not apparent. The typical noise is the median of all these spaxels (21 spaxels in total), $2.36\times10^{-18} \pm 1.1\times10^{-18}$\flux. The middle panels of \autoref{fig:sim} show the flux map including the clumps (left panel) and the velocity and velocity dispersions (middle and right panel) recovered with \pan~after adding noise. 

3) We convolve the simulated galaxy by the OSIRIS AO PSF. This AO PSF was measured from the telluric stars observed each night. We found that the median AO PSF is composed by a Gaussian core $FWHM=1.39\pm0.74$ spaxels ($\sim0.\arc14$) and side wings that can be described by a second Gaussian $FWHM=7 \pm 1.0$ spaxels \citep[see][for details on the AO PSF and its comparison with other observations]{Bassett_2016}. We test the effects of the AO PSF in the simulation using the actual profile that describes:

a) the median AO PSF; b) the widest AO PSF in our observations (Gaussian core $FWHM=3\pm0.51$ spaxels and side wings $FWHM=7 \pm 1.0$ spaxels). The bottom panels of \autoref{fig:sim} show flux and kinematic maps (recovered from \pan) after convolving the galaxy with the widest AO PSF (case b). 
We can see the AO PSF effect at the centre of the galaxy and cross-like patterns created by the noise. 

\begin{figure}
 \includegraphics[width=\linewidth]{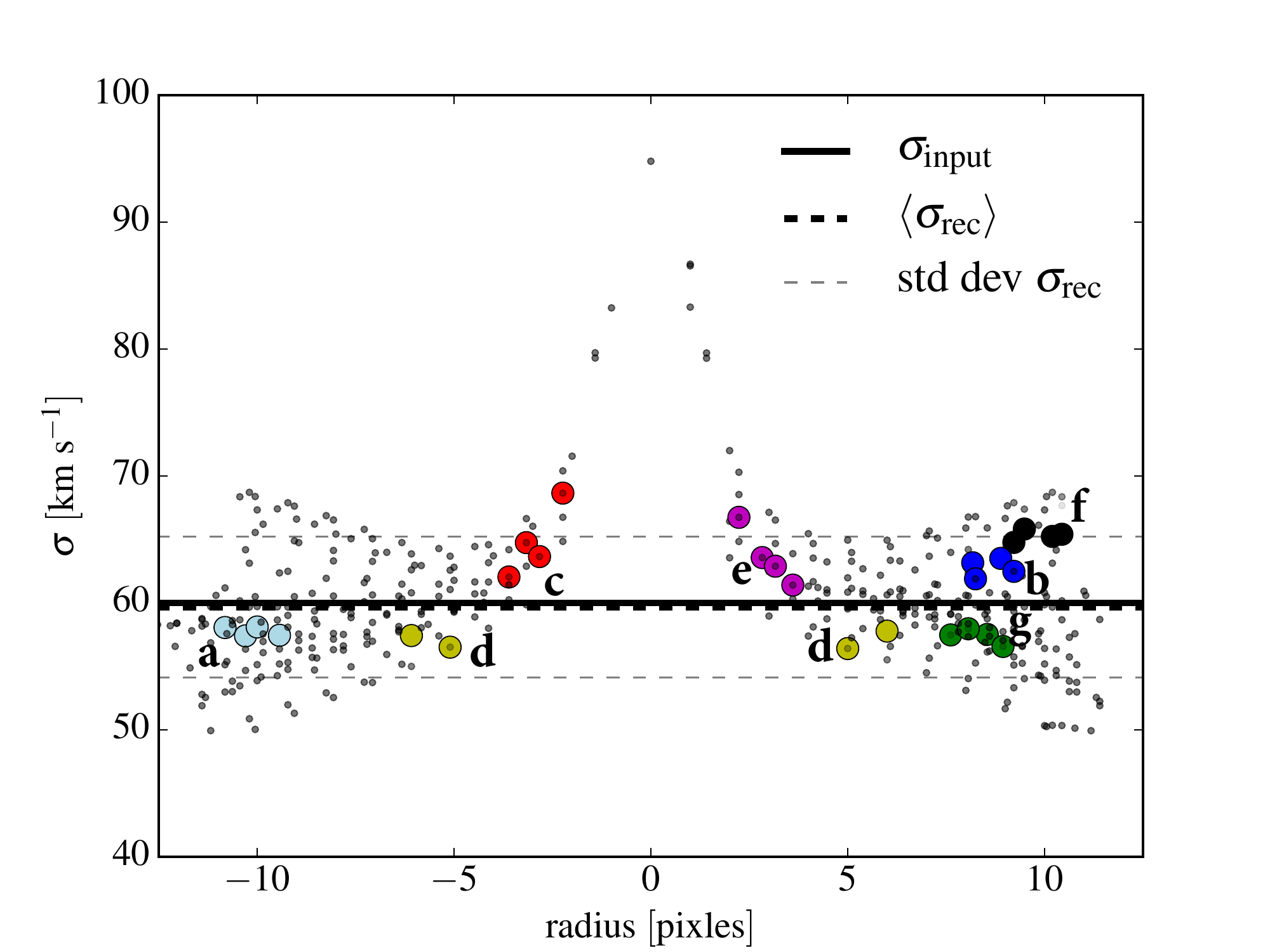}
 \caption{Results from the AO PSF and noise simulations. This figure shows the velocity dispersion of each spaxel in the simulated galaxy (from \autoref{fig:sim}) as a function of their radial distance to the galactic centre. The black dots represent the spaxels in the galaxy floor. The coloured circles represent the spaxels within clumps. The labels (lower case letters) correspond to the clumps in bottom-left panel in \autoref{fig:sim}. The solid line shows the input velocity dispersion (60 \kms), the thick dashed line shows the recovered median velocity dispersion (59.7 \kms) after adding the noise and AO PSF effects to the galaxy. The thin dashed lines are the standard deviations (5.5 \kms). The velocity dispersion is best recovered in the star-forming clumps (mean $\sigma_{\rm clumps}=60.5\pm2.2$~\kms) due to their high S/N. A colour version of this plot is available at the electronic edition.
	}
\label{fig:sim_res}
\end{figure}

The AO PSF mostly affects the centre of the galaxies (cases a and b), elevating the velocity dispersion more than two standard deviations. When using the widest AO PSF (case~b) it can also elevate the velocity dispersion in regions between clumps. However, the rise in velocity dispersion in these regions would be of the order of one standard deviation (clump `f': black circles in \autoref{fig:sim_res}), which is within the measurement errors.  
\autoref{fig:sim_res} shows the velocity dispersion of each spaxel in the simulated galaxy (from \autoref{fig:sim}) as a function of their radial distance from the centre. The coloured circles represent the spaxels within the star-forming clumps. We labelled the clumps with lower case letters to locate them in the bottom-left panel of \autoref{fig:sim} and in \autoref{fig:sim_res}. The solid line represents the input velocity dispersion (60~\kms) and the dashed lines represent the median (59.7~\kms; thick line) and standard deviation (5.5~\kms; thin lines) of the velocity dispersion recovered from \pan. Spaxels coincident with star-forming regions best recover the input velocity dispersion (mean $\sigma_{\rm clumps}=60.5$~\kms standard deviation 2.2~\kms), except for those within 3 spaxels of the galactic centre. The relatively higher S/N in the clumps clearly drives the more accurate velocity dispersion measurement. As we reach the edge of the galaxy, the S/N decreases and the systematic effect increases. 
\\\\
This process of creating models and measuring velocity dispersions is repeated 100 times with different clump configurations, i.e. random number of clumps, sizes, positions, and brightness. From the results of the simulation we fit the uncertainty in velocity dispersion ($\Delta \sigma$) due to AO PSF and noise effects, as a function of the S/N:
\begin{equation}
\Delta \sigma = 11.486 \exp{(-0.1\times \rm S/N)}+0.442
\end{equation}

We find that if we apply a S/N cut of 4 all spaxels satisfy $\Delta \sigma<10$\kms. Therefore, we masked all spaxels with S/N~$<4$. We also exclude from our analysis clumps within 3 spaxels of the galactic centre, as they are highly affected by spatial sampling and possible AGN contamination. With this function we create a systematic map per galaxy, to later subtract it from the measured velocity dispersion maps. This reduces the AO PSF and noise systematic effects, and we can reliably compare the velocity dispersion from the inner regions of the galaxy (excluding $\pm 3$ from the galactic centre). 

\subsection{Kinematic modeling}\label{sec:mod}
\begin{figure}
     \includegraphics[width=0.9\columnwidth]{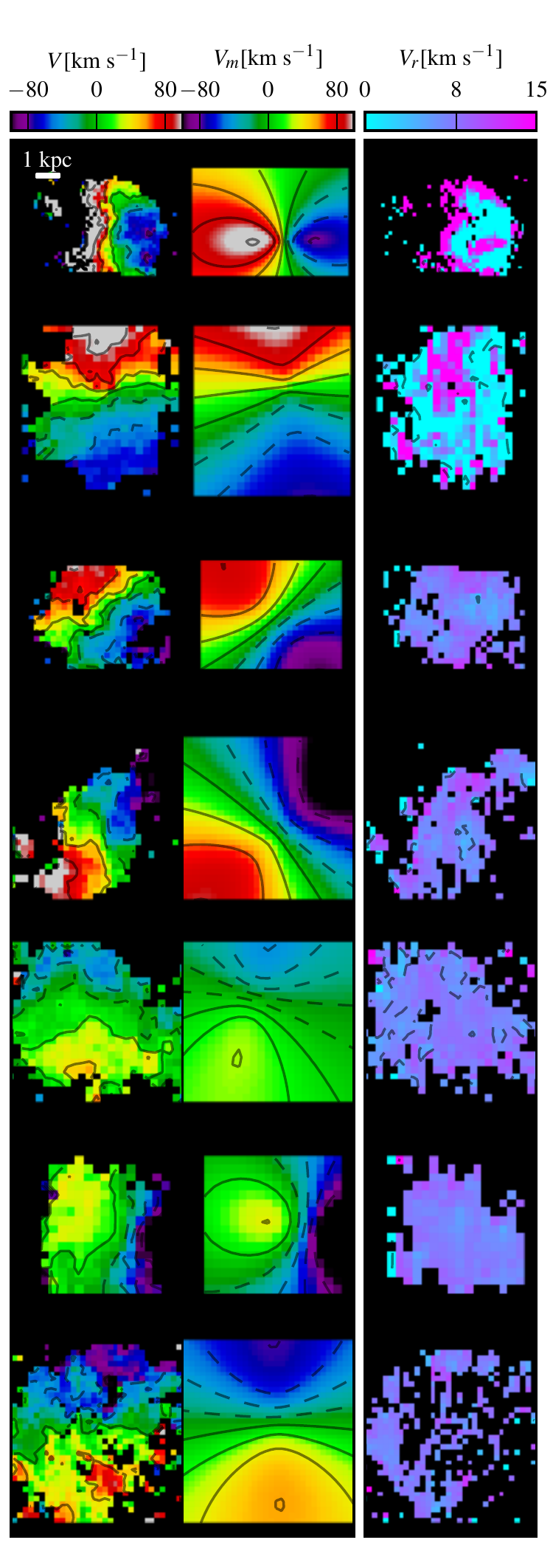}
    \caption{Velocity maps. The \textbf{left-hand panels} show the observed velocities, the \textbf{middle panels} show the rotation curve fitted with \autoref{eq:epinat} \citep{Epinat_2008a} and the \textbf{right-hand panels} show the residuals of the fits. The name of each galaxy is at the top of the right-hand panels.  The images are shown at the same physical scale to facilitate visual inspection (top left corner). A colour version of this plot is available at the electronic edition. 
    	}
    \label{fig:model}
\end{figure}
\begin{figure}
     \includegraphics[width=0.9\linewidth]{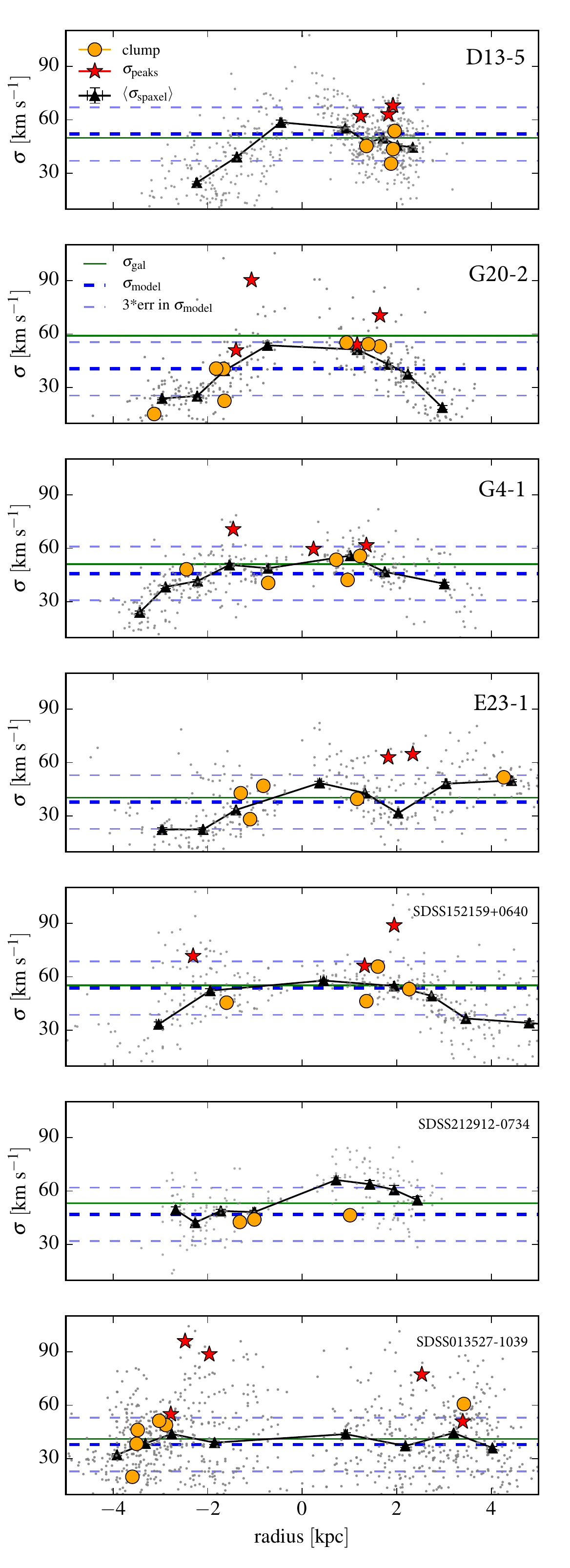}
    \caption{Velocity dispersion as function of the radial distance from the centre of the galaxy. The black dots represent each spaxel. The black triangles show the median velocity dispersion of each distance bin (each bin has the same number of spaxels). The orange circles mark the brightest spaxel of each clump. The red stars are the \sigp~described in \autoref{sec:peaks}. The name of each galaxy is at the right-top corner. We show the constant $\sigma_{\rm model}$ and 3$\times$ the error in $\sigma_{\rm model}$ in dashed blue lines. The luminosity weighted velocity dispersion of the galaxy is shown as a green solid line. There is statistically significant structure in 6 out 7 galaxies. A colour version of this plot is available at the electronic edition.
    	}
    \label{fig:vd_profiles}
\end{figure}
We fit a disc model to our galaxies to apply a spatial sampling correction and to compare the differences in velocity and velocity dispersion maps between models and observations. We use the automatized tool \gbk\footnote{\url{http://supercomputing.swin.edu.au/gbkfit}}~\citep{Bekiaris_2016}, a software designed to model the kinematics of IFS data. \gbk~uses parallel architectures such as multi-core CPUs and Graphic Processing Units (GPU), to model thousands of galaxies in short time frames. 

\gbk~was tested with large data sets such as the galaxies in the SAMI survey \citep{Bryant_2015}, the GHASP survey \citep{Epinat_2008a} and the first data set of the DYNAMO galaxies \citep[observed with SPIRAL and WiFES;][]{Green_2014}. \gbk~has accurately recovered previous modelling results of these IFS observations.  

The software uses the velocity and velocity dispersion maps measured from the observations to fit a rotating disc model with an intrinsic velocity dispersion ($\sigma_{\rm model}$), constant across the galaxy \citep{Davies_2011}. The correction for spatial sampling is achieved by convolving the modelled galaxy with a 3-dimensional function that combines the line and point spread function \citep[instrument spectral resolution and PSF; Equation 9 in][]{Bekiaris_2016}. 

The model fitting method is presented in \citet{Bekiaris_2016} and consist in a) evaluate the observed cube using input parameters; b) convolve the model cube with the line and point spread function of the observations; c) evaluate the goodness of fit by calculating the $\chi^2$ of the fit from the residuals; d) evaluate when the termination criteria is met. If not, the process begins again changing the initial parameters. 

We ran the following rotation curve models for the DYNAMO-OSIRIS sample: flat, arctan \citep{Courteau_1997}, and the analytical function of \citet{Epinat_2008a}; with the optimisers: Levenberg-Marquardt \citep{Barbosa_2006, Coccato_2008} and Nested Sampling \citep{Skilling_2004}. We also tried different priors for the galactic centre: fixed centre, allowing to vary $\pm2$ spaxels from the initial conditions, and allowing to vary $\pm10$ spaxels from the initial conditions. 

We find that a Levenberg-Marquardt optimiser, a flexible galactic centre within $\pm2$ spaxels, and the analytical function of \citet{Epinat_2008a} give the best goodness of fit for our galaxies. The four-parameter analytical function of \citet{Epinat_2008a} is given by:
\begin{equation}
V_{\text{rot}} = V_t \frac{(r/r_t)^g}{1+(r/r_t)^a}
\label{eq:epinat}
\end{equation}
Where $V_{\text{rot}}$ is the rotation curve, $V_t$ and $r_t$ are the turn-over velocity and radius. The constants $g$ and $a$ control the inner and outer slope of the rotation curve. The rotation curve is flat if $a=g$, rising if $a<g$, and declining if $a>g$.

We calculate the goodness of fit, $\chi^2$, from Equation 11 in \citet{Bekiaris_2016}, which adds in quadrature the residuals from the velocity map with the weighted ($W$) residuals from the velocity dispersion map. We tried $W=1$ and $W=0.2$. However, we chose to use $W=0.2$ as \gbk~fits a constant velocity dispersion across the galaxy, and this does not always represent the velocity dispersion in our galaxies. The reduced $\chi^2$ from the fit of our sample span from 0.31 to 2.26 (\autoref{tab:gal_prop}), representing a reliable fit. \autoref{fig:model} shows the observed velocity map (left-hand panels), the modelled velocity map (middle panels), and the residuals from the fit (right-hand panels). These maps show that the analytical function of \citet{Epinat_2008a} fits the observable data $\pm15$ \kms. 

We calculate the shear in the observed (\vs) and modelled velocity $V_{\rm s,mod}$ as ($V_{\rm max}$ -$V_{\rm min} $)/2 where $V_{\rm max}$ and $V_{\rm min}$ are the means of the spaxels that contain the 5\% highest and lowest velocities in the masked region \citep{Law_2009,Wisnioski_2012}. We do not correct the \vs~from galaxy inclination because we do not rely on the inclinations we have for the DYNAMO-OSIRIS galaxies. The inclinations were taken from the galaxy SDSS photometry and are used only as a proxy in the rotation curve fitting. We expect the velocities to increase with lower inclinations as $V_{\rm shear} \sim V_{\rm l.o.s}/sin\text{ }i$. i.e. If the SDSS inclinations were accurate the \vs~of SDSS013527-1039 would increase by 64\% and the \vs~of D13-5 would increase by 26\%.

We use the modelled velocity maps to calculate the velocity gradient per spaxel and apply a spatial sampling correction to our observed velocity dispersion maps \citep{Davies_2011, Varidel_2016}. The velocity gradient in each spaxel is subtracted in quadrature from the measured velocity dispersion. 

The final observed velocity dispersion profiles (black triangles) and the $\sigma_{\rm model}$ (dashed blue thick line) are shown in \autoref{fig:vd_profiles}. The observed $\sigma_{\rm gal}$ is calculated from de-rotating the galaxy and stacking all the spectra within the masked region (luminosity weighted velocity dispersion; green solid line). Different methods have been proposed in the literature \citep[for a review]{Davies_2011} to calculate $\sigma_{\rm gal}$. The global properties of the galaxies does not lie within the scope if this study, and thus the chosen method does not affect our results. \autoref{tab:gal_prop} summarizes the galaxy properties \vs, $\sigma_{\rm gal}$, $V_{\rm s,mod}$, $\sigma_{\rm model}$, \textit{incl}, and reduced $\chi^2$. \textit{incl} is the inclination of the disc from SDSS photometry. 

\begin{table*}
	\centering
	\caption{Galaxy properties. Column 1 shows the DYNAMO or SDSS ID of the galaxy. Column 2 presents the number of clumps identified per galaxy. Column 3 shows the dust corrected \lpaa~measured within a $\sim3\arc$ aperture from the OSIRIS observations. Column 4 shows the \lpaa~uncertainties from flux calibration errors. Column 5 shows the luminosity weighted velocity dispersion.  Column 6 presents the measured velocity shear within the OSIRIS coverage of the galaxy. The uncertainties in Column 5 and 6 where measured using the Monte-Carlo modelling in \pan, which shuffles the spectral errors and repeats the fit 100 times. Column 7, and 8 show the kinematic properties of the galaxy returned by the analytical function of \citet[][\autoref{sec:mod}]{Epinat_2008a}. Column 9 is the inclination of the projected disc from SDSS photometry. Column 10 shows the reduced $\chi^2$ from each fit. }
	\label{tab:gal_prop}
	\begin{tabular}{lccccccccccc} 
		\hline
		ID &N of & \lpaa$_{\rm corr}$ &\lpaa$_{\rm corr}$~err& $\sigma_{\rm gal}$ & \vs&$\sigma_{\rm model}$&$V_{\rm s,mod}$&incl&reduced\\
		             &clumps&[erg s$^{-1}]$&[erg s$^{-1}]$&$\pm7$ [\kms]&$\pm 10$ [\kms]&$\pm5$[\kms]&$\pm7$[\kms]&[deg]&$\chi^2$\\
		\hline
		  D13-5&4&$1.14\times10^{41}$&$2.28\times10^{40}$&49.85&163.32&51.95&150.14&48.65&1.25\\ 
		  G20-2&7&$1.50\times10^{41}$&$3.45\times10^{40}$&58.97&104.48&40.49&138.72&33.98&0.54\\
		  G4-1&5&$1.50\times10^{41}$&$3.90\times10^{40}$&51.04&134.31&45.88&131.19&23.99&0.51\\
		  E23-1&5&$2.87\times10^{41}$&$5.74\times10^{40}$&40.27&156.94&37.88&151.72&44.27&2.26\\
		  SDSS152159+0640&4&$1.53\times10^{42}$&$3.21\times10^{41}$&55.26&88.11&53.63&50.06&35.92&0.56\\
		  SDSS212912-0734&3&$1.01\times10^{42}$&$3.03\times10^{41}$&53.16&102.41&46.93&105.94&32.32&0.37\\
		  SDSS013527-1039&6&$5.34\times10^{41}$&$1.21\times10^{41}$&41.17&179.91&38.01&94.20&21.01&0.31\\
		\hline
	\end{tabular}
\end{table*}

\section{Results}

The first objective of this work is to confirm significant structure in the velocity dispersion maps. The second, is to study the associate velocity dispersions of star-forming clumps. 

We find that all 7 galaxies are sufficiently consistent with kinematic disc models, and it is therefore reasonable to classify them as discs. The velocity dispersion maps have been corrected for spatial sampling, noise and AO PSF systematics.  We ignore clumps near the galactic centre ($\pm3$ spaxels), as this region can be contaminated by AGN activity and it is the most affected by systematics. 
\subsection{Peaks in the velocity dispersion maps}\label{sec:peaks}
\begin{figure*}
     \includegraphics[width=1\linewidth]{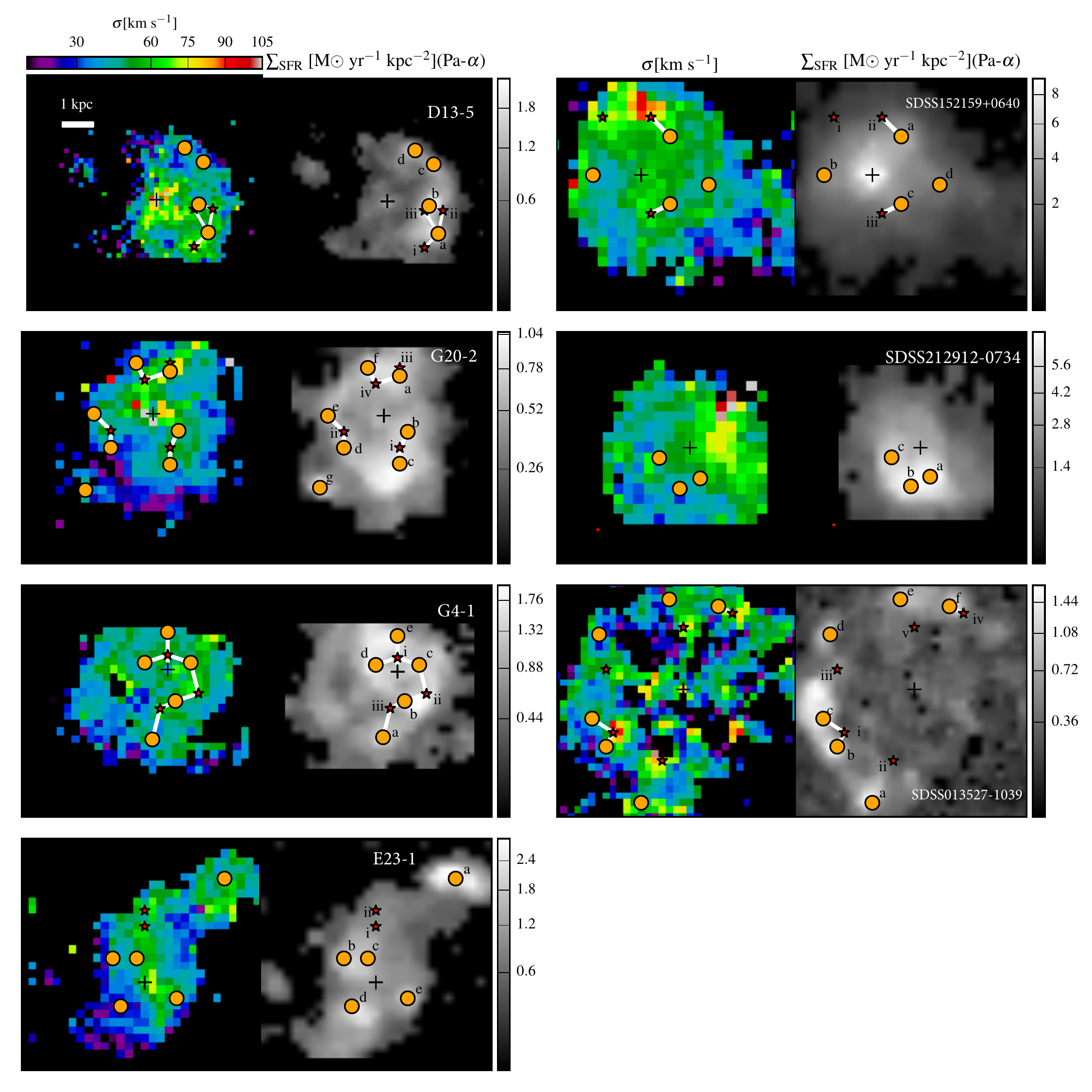}
    \caption{Velocity dispersion (left-hand panels) and \paa~(right-hand panels) maps. The name of each galaxy is at the right-top corner of the \paa~maps. We show the \sigp~as red stars labelled by Roman numbers, and the brightest spaxel of the star-forming clump (clump centre) as orange circles labelled by lowercase letters. The crosses show the centre of each galaxy. The images are shown at the same physical scale to facilitate visual inspection (top left corner of D13-5). The white lines connect the star-forming clumps and \sigp~within 1~kpc away (associated clumps). $\sim67$\% of the \sigp~suggest a connection to star-forming clumps. A colour version of this plot is available at the electronic edition.
    	}
    \label{fig:vd_paa}
\end{figure*}

\begin{figure*}
     \includegraphics[width=1\linewidth]{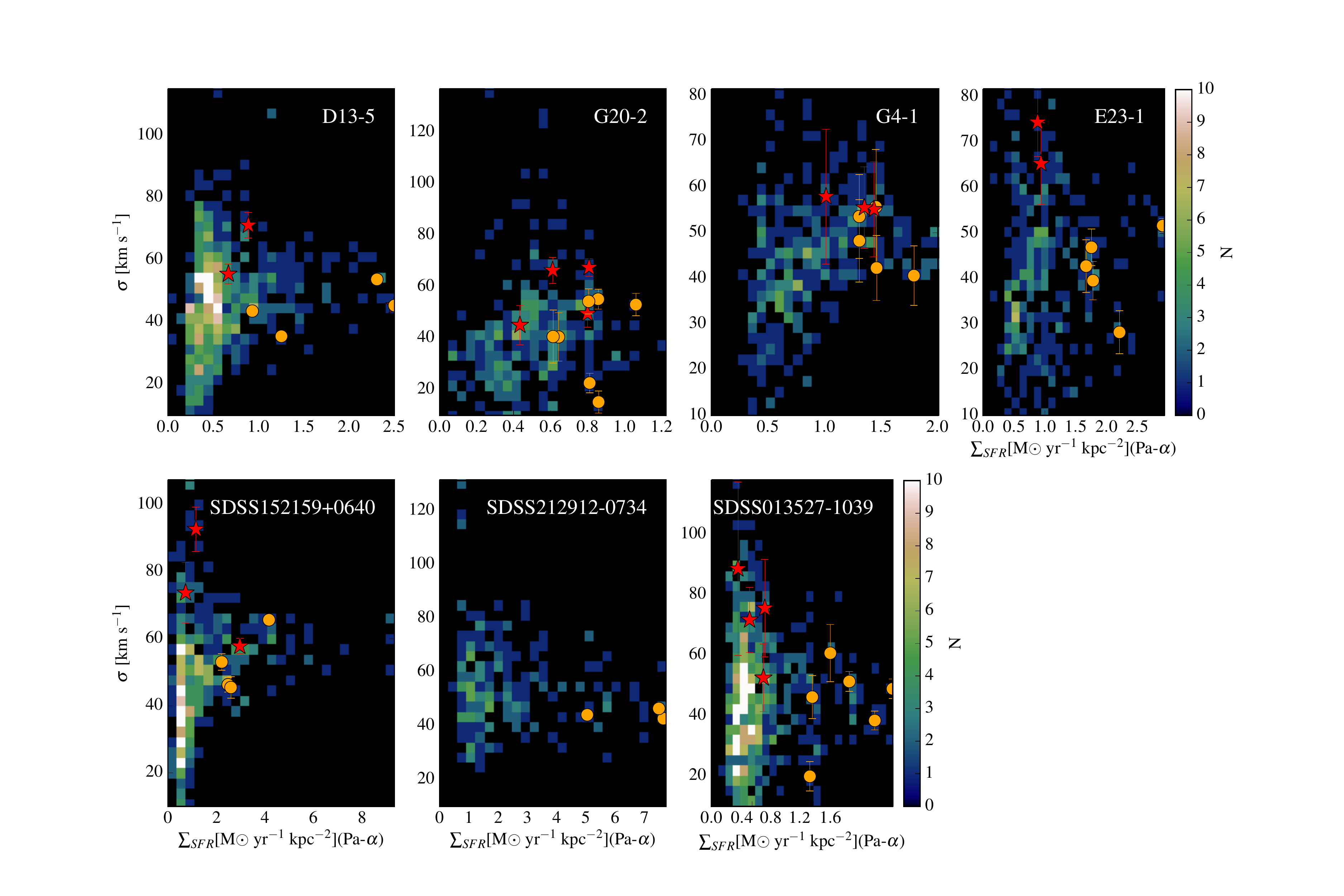}
    \caption{Velocity dispersion as a function of \sigsfr~(per spaxel). The name of each galaxy is at the top of each panel. We show the \sigp~as red stars and the brightest spaxel of the star-forming clump (clump centre) as orange circles. In G20-2 and G4-1 the velocity dispersion increases with increasing \sigsfr. While the rest of the galaxies do not show a clear trend. A colour version of this plot is available at the electronic edition.
    	}
    \label{fig:vd_sfr}
\end{figure*}

\begin{figure}
     \includegraphics[width=1\linewidth]{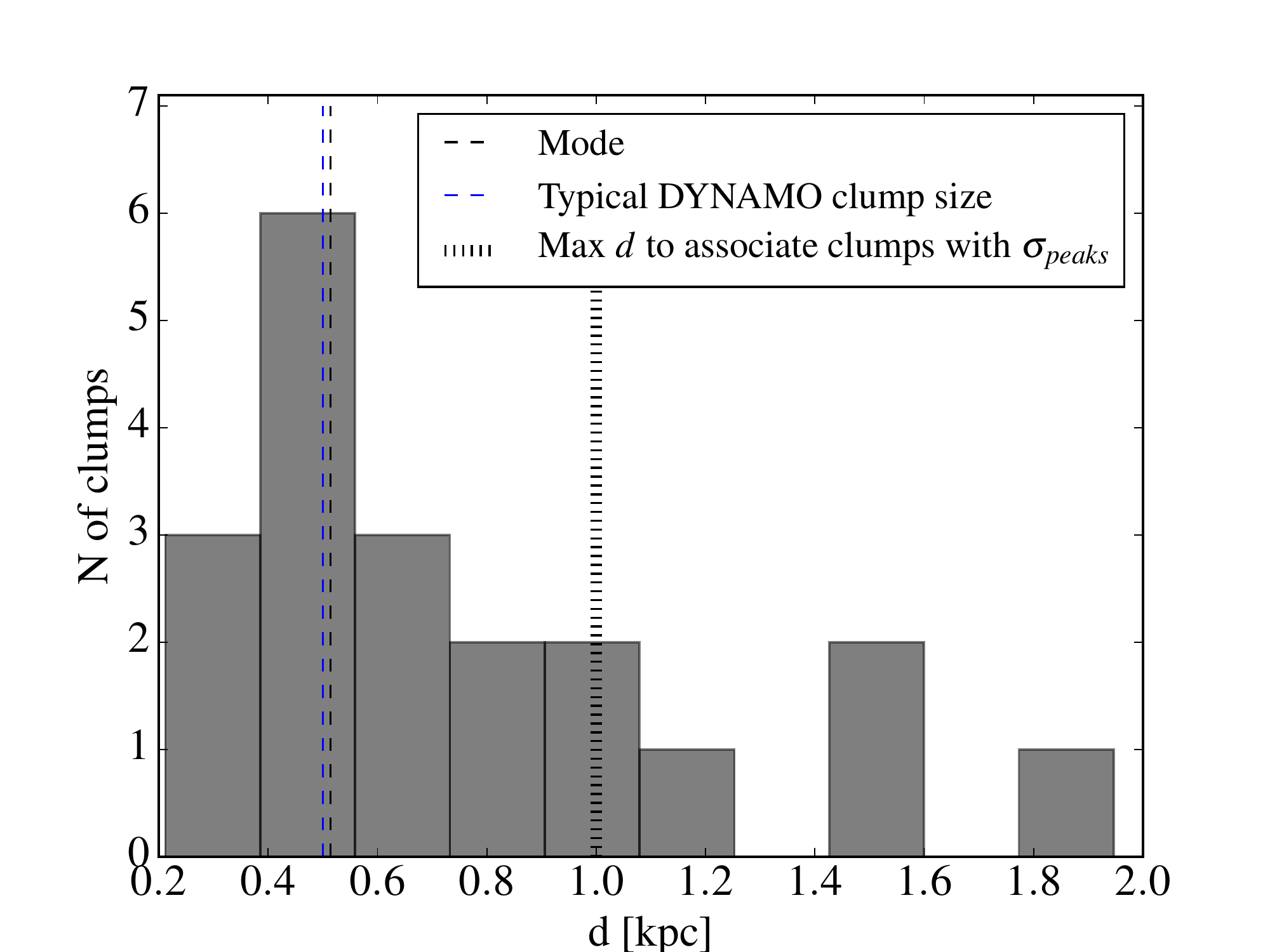}
    \caption{Distance to the nearest clump. The histogram shows the distribution of the distance from the \sigp~to the nearest star-forming clump. 
    The black dashed line shows the mode of the distribution. The blue dashed line shows the typical size of DYNAMO clumps \citep{Fisher_2016}. The dotted line shows the maximum distance at which we would expect a connection between a star-forming clump and a \sigp. The majority of the \sigp~($\sim67$\%) appear to be associated to star-forming clumps.}
    \label{fig:dist}
\end{figure}
\begin{figure}
     \includegraphics[width=1\linewidth]{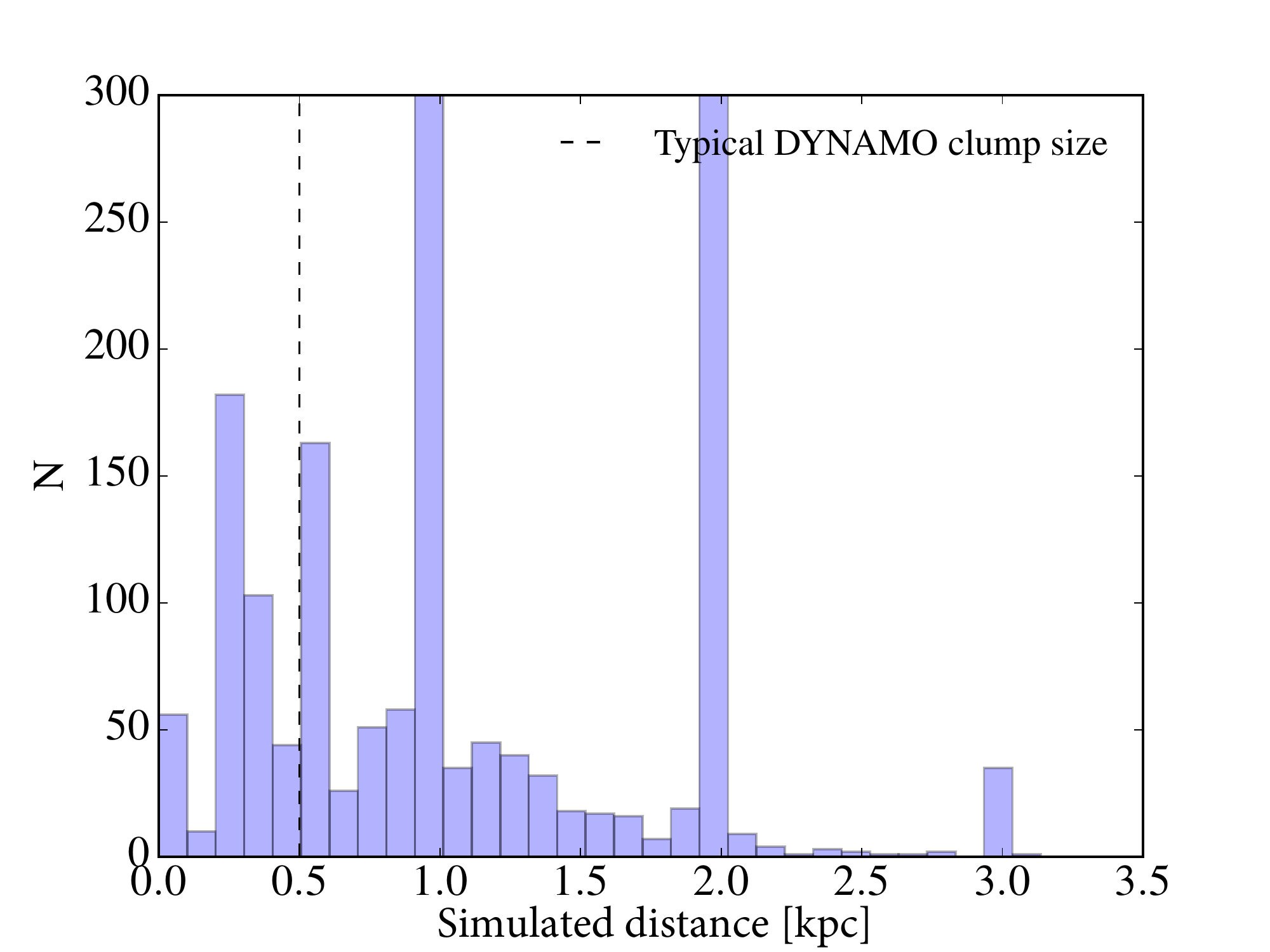}
    \caption{Modes of 1000 distributions of distances from \sigp~to the nearest clump in simulated galaxies. 
    The dashed line shows the typical size of DYNAMO clumps \citep{Fisher_2016}.}
    \label{fig:ran_dist}
\end{figure}

We identify the peaks of the distribution in the velocity dispersion maps, to study the velocity dispersion structure. We call \textit{peaks} (\sigp) the spaxels with velocity dispersion higher or equal to the velocity dispersion cut $\sigma_{\rm cut} = \sigma_{\rm model} + 3\times \text{error in } \sigma_{\rm model}$. Where $\sigma_{\rm model}$ is the velocity dispersion fitted by the analytical function of \citet{Epinat_2008a} with \gbk; The $\langle 3\times \text{error in } \sigma_{\rm model}\rangle \sim15$\kms. We do not take isolated spaxels, only peaks which comprise 3 or more spaxels with velocity dispersion $\geq \sigma_{\rm cut}$. This is still visible after smoothing the galaxy with a $0.\arc3$ Gaussian.

As we mention earlier, the instrumental resolution of OSIRIS varies across the FOV. We therefore evaluate the impact of this on our results. We find that the instrumental resolution in our observations varies from 35 to 47 \kms. This variation occurs smoothly across the chip and it is small compared to $\sigma_{\rm cut}$. We find that the variation in the instrumental resolution is it not the source of the \sigp. We also find that none of the selected \sigp~lie in zones of low instrumental resolution. We tested regions equivalent to the size of a clump ($\sim 0.\arc7$) and we find that the maximum variation in such areas is 3\kms. However, we do not fully discard the effect of this variation and we account for it by adding it in quadrature to the measurement errors.

\autoref{fig:vd_profiles} shows the \sigp~that pass this criteria (red stars), $\sigma_{\rm model}$ is shown as a thick blue dashed line and the $\langle 3\times \text{error in } \sigma_{\rm model}\rangle$ are shown as thin blue dashed lines. On average the velocity dispersion in \sigp~across the sample is $66.07\pm12$~\kms. This is 18~\kms~higher than the average velocity dispersion of the galaxy floor across the sample, $48\pm8$ \kms.

In 6 out of 7 galaxies we find $2-5$ velocity dispersion peaks. We can, therefore, confirm significant structure in the velocity dispersion of turbulent galaxies. There is only one galaxy, SDSS212912-0734, which does not show significant structure in the velocity dispersion map (median $\sigma_{\rm spaxel} = 53.07\pm10.19$\kms), the areas of relatively elevated velocity dispersion are close to the galactic centre. A summary of the \sigp~per galaxy can be found in \autoref{tab:sigp}.

\begin{table*}
	\centering
	\caption{\sigp~properties. Column 1 shows the DYNAMO or SDSS ID of the galaxy. Column 2 show the \sigp~IDs. Column 3 and 4 present the \sigp~values. The uncertainties come from 100 Monte Carlo realisation of the emission line fit. Column 5 shows the ID of the associated star-forming clump. Column 6 shows the distance to the nearest clump (\autoref{fig:dist}). }
	\label{tab:sigp}
	\begin{tabular}{lccccc} 
		\hline
		ID &\sigp&\sigp&\sigp~err&assoc to&d\\
		             &ID&[\kms]&[\kms]&clump&[kpc]\\
		\hline\hline
		  D13-5&i&67.91&8.01&a&0.63\\
D13-5&ii&63.02&5.78&a,b&0.47\\
D13-5&iii&62.04&2.40&a,b&0.21\\
\hline
G20-2&i&54.02&9.53&b,c&0.52\\
G20-2&ii&49.66&14.99&d,e&0.52\\
G20-2&iii&70.42&7.60a&0.26\\
G20-2&iv&90.18&8.60&a,f&0.58\\
\hline
G4-1&i&59.52&21.45&c,d,e&0.72\\
G4-1&ii&61.65&23.50&c,b&0.75\\
G4-1&iii&70.63&25.10&a,b&0.53\\
\hline
E23-1&i&62.92&16.01&none&1.07\\
E23-1&ii&64.70&25.60&none&1.58\\
\hline
SDSS152159+0640&i&71.61&20.10&none&1.94\\
SDSS152159+0640&ii&88.78&9.50&a&0.90\\
SDSS152159+0640&iii&66.09&5.64&c&0.71\\
\hline
SDSS212912-0734&none&none&none&none&none\\
\hline
SDSS013527-1039&i&95.88&34.23&b,c&0.51\\
SDSS013527-1039&ii&88.56&25.60&none&1.54\\
SDSS013527-1039&iii&54.99&22.45&none&1.17\\
SDSS013527-1039&iv&50.87&12.62&f&0.51\\
SDSS013527-1039&v&77.22&22.56&none&1.02\\
		\hline
	\end{tabular}
\end{table*}

\subsection{Are the star-forming clumps connected to the velocity dispersion structure?}
\autoref{fig:vd_paa} shows the velocity dispersion maps (left-hand panels) and \sigsfr~maps (right-hand panels) of each galaxy. The brightest spaxel of each star-forming clump is marked with a lowercase letter and an orange circle. The \sigp~are marked with lowercase roman numbers and red stars. 

We analyse whether there is or is not a direct correspondence between $\sigma$ and \sigsfr~in each galaxy, by plotting the $\sigma$ vs \sigsfr~(per spaxel; \autoref{fig:vd_sfr}). We show the \sigp~as red stars and the star-forming clumps as yellow circles. Note that the \sigp~are not always the spaxels with highest velocity dispersion. This is because in most of the galaxies the highest velocity dispersions are found in the centre of the galaxy. 

 G20-2 and G4-1 show an increase in velocity dispersion with increasing star formation rate density. This is not the case for the rest of the galaxies in our sample. We do not see any correspondence between $\sigma$ and \sigsfr~in D13-5, G4-1, E23-1, SDSS152159+0640, SDSS212912-0734, and SDSS013527-1039. We also explored the velocity dispersion in each star-forming clump by plotting $\sigma$ as a function of the spaxel distance to the clump centre. We do not see any trend in any clump. 
 
We examine the distances between \sigp~and the star-forming clumps. \autoref{fig:dist} shows the distribution of the distances\footnote{The distances are not corrected by galaxy inclination because the inclinations were taken from SDSS photometry and we do not completely rely on them.} from the \sigp~to the nearest clump. The mode of the distribution is 0.52~kpc. We note that this is very similar to the average radius encompassing 90\% of the clump flux of clumps in DYNAMO galaxies \citep{Fisher_2016,Fisher2017}. We find that 67\% of \sigp~are located within 1~kpc of a clump centre identified in \paa~emission line flux. Henceforth, we refer to \sigp~sitting at less than 1~kpc away from the clump centre as \textit{associated} \sigp, suggesting a possible link between the star-forming clumps and the \sigp. In \autoref{fig:vd_paa} we present this \textit{association} as white lines.

To rule out any bias in the distance to the nearest clump distribution, we simulate each galaxy by fixing the \sigp~and moving the clumps to random positions. Then we measure the distances from the \sigp~to the each clump and reproduce \autoref{fig:dist} (distribution of the distances from the \sigp~to the nearest clump). We repeat this procedure 1000 times varying the clump sizes. \autoref{fig:ran_dist} shows the modes of the 1000 runs. 

The distances from each \sigp~to the nearest clump peak at 2.0, 1.0, 0.51 and 0.25 kpc. Although, the distribution of the random distances is different from the distribution of the distances from \sigp~to the nearest clump in the observations (the probability of the two samples to be drawn from the same distribution is 1.02\%, Kolmogorov-Smirnov test), in 155/1000 simulations the typical distances is 0.51~kpc similar to what we see in our observations (0.52~kpc). Therefore, $\sim15$\% of the simulations suggest that the typical distance observed between a \sigp~and the closest clump is due to systematics, i.e. limited coverage of the galaxies by the OSIRIS FOV.  Observations covering more than 7~kpc of a galaxy will be needed to confirm our result, that \sigp~sit within 1~kpc of clump centres. 

It is important not to confuse the association between star-forming clumps and \sigp~with a physical correlation. Not all star-forming clumps have associated \sigp~and vice-versa. e.g. SDSS212912-0734 does not have any \sigp~and the \sigp~in E23-1 do not appear to be associated with the star-forming clumps. We would like to stress that this `lack of correlation' can also be biased by the small coverage of our observations. At least 44\% of the clumps sit less than 1kpc away of the edge of the OSIRIS galaxy coverage. Therefore, there is a possibility that more \sigp~are lying $\sim1$~kpc outside the coverage of our observations. The summary of the star-forming clumps properties is in \autoref{tab:clumps}. 
\begin{table*}
	\centering
	\caption{Star-forming clump properties. Column 1 shows the DYNAMO or SDSS ID of the galaxy. Column 2 shows the clump ID. Column 3 and 4 present the \paa~Luminosity of the brightest spaxel in the clump. Uncertainties come from flux calibration. Column 5 and 6 present the velocity dispersion of the brightest spaxel in the clump. The uncertainties come from 100 Monte Carlo realisations of the emission line fit. Column 7 shows the ID of the associated \sigp.}
	\label{tab:clumps}
	\begin{tabular}{lcccccc} 
		\hline
	 ID &clump&\lpaa$_{\rm corr}$&\lpaa$_{\rm corr}$~err&$\sigma_{\rm clump}$&$\sigma_{\rm clump}$ err&assoc to\\
		             &ID&[$10^{39}$ erg s$^{-1}$]&[$10^{39}$ erg s$^{-1}$]&[\kms]&[\kms]&\sigp\\
		\hline\hline
D13-5&a&1.28&0.25&53.71&2.75&i,ii, iii\\
D13-5&b&1.39&0.27&45.31&1.81&ii, iii\\
D13-5&c&0.52&0.10&43.55&3.65&none\\
D13-5&d&0.70&0.14&35.44&2.75&none\\
\hline
G20-2&a&1.34&0.26&54.31&10.74&iii, iv\\
G20-2&b&1.43&0.28&55.17&8.53&i\\
G20-2&c&1.77&0.35&53.05&9.63&i\\
G20-2&d&1.07&0.21&40.50&20.68&ii\\
G20-2&e&1.02&0.20&40.64&22.72&ii\\
G20-2&f&1.35&0.27&22.59&8.32&iv\\
G20-2&g&1.43&0.28&15.20&9.43&none\\
\hline
G4-1&a&1.86&0.37&48.24&19.84&iii\\
G4-1&b&2.07&0.41&55.66&27.62&ii, iii\\
G4-1&c&1.86&0.37&53.56&20.17&i,ii\\
G4-1&d&2.55&0.51&40.61&14.33&i\\
G4-1&e&2.08&0.41&42.28&15.65&i\\
\hline
E23-1&a&4.86&0.97&51.71&3.46&none\\
E23-1&b&2.78&0.55&42.80&12.75&none\\
E23-1&c&2.93&0.58&46.95&8.87&none\\
E23-1&d&3.69&0.73&28.27&10.41&none\\
E23-1&e&2.97&0.59&39.62&9.29&none\\
\hline
SDSS152159+0640&a&10.53&2.1&65.68&3.05&ii\\
SDSS152159+0640&b&6.56&1.31&45.50&7.07&none\\
SDSS152159+0640&c&6.26&1.25&46.33&4.81&iii\\
SDSS152159+0640&d&5.60&1.12&53.10&5.42&none\\
\hline
SDSS212912-0734&a&18.96&3.79&46.43&2.93&none\\
SDSS212912-0734&b&19.35&3.87&42.65&2.91&none\\
SDSS212912-0734&c&12.76&2.55&44.06&3.36&none\\
\hline
SDSS013527-1039&a&2.09&0.41&38.31&6.83&none\\
SDSS013527-1039&b&2.32&0.46&48.90&7.12&i\\
SDSS013527-1039&c&1.77&0.35&51.32&7.25&i\\
SDSS013527-1039&d&1.26&0.25&19.94&10.64&none\\
SDSS013527-1039&e&1.29&0.25&46.16&15.58&none\\
SDSS013527-1039&f&1.52&0.30&60.69&20.81&iv\\
		\hline
	\end{tabular}
\end{table*}

\section{Discussion}
In this paper we analyze the kinematics of a sample of clumpy, turbulent discs at $z\sim0.1$ with spatial sampling of 150 to 400~pc. This type of galaxy is rare in the local Universe, but they represent $\sim1/3$ of the high star-forming galaxy population at high redshifts.   
Their properties (high SFR, young stars) suggest a rapid formation \citep[$<1$~Gyr;][]{Genzel_2006, Forster_2006} and their kinematics reflect a fast, however, smooth accretion process \citep[e.g.][]{Shapiro_2008}.

We find significant structure in the velocity dispersion of the DYNAMO-OSIRIS turbulent discs. 6 out of 7 galaxies in our sample have \sigp~with velocity dispersion higher than the velocity dispersion of the galaxy floor ($\sigma_{\rm peaks} \geq \sigma_{\rm model} + 3\times \text{error in } \sigma_{\rm model}$; \autoref{fig:vd_profiles}). This confirms the weak structure seen in high redshift observations at $\sim3\times$ lower spatial resolution \citep{Genzel_2011,Wisnioski_2012, Newman_2012a, Newman_2012b}.  

A clump can lose angular momentum due to clump-clump encounters, dynamical friction, gas drag, mass exchange with the disc, etc. \citet{Ceverino_2012} predicted that some of the main sources of support against this gravitational collapse in clump evolution are turbulence and rotation. \citet{Ceverino_2012} assumes that the velocity dispersion can be driven by turbulence. They presented a case where rotation provides $\sim50$\% of support, and the velocity dispersion ($\sigma_{\rm clump}$) is comparable to the velocity dispersion of the disc. We find that most of the clumps in the DYNAMO-OSIRIS galaxies have $\sigma_{\rm clump} = \sigma_{\rm model} \pm 15$~\kms. Consistent with \citet{Ceverino_2012}'s prediction. This would suggest that velocity dispersion (turbulence) is providing some of the support against gravitational collapse, being complemented by some other mechanism (e.g. rotation).

We find that 67\% of the \sigp~in our galaxies lie within 1~kpc of a clump centre (most common distance 0.5~kpc). Suggesting that \sigp~can be found at the edges of the star-forming clumps (\autoref{fig:dist}). The position of the \sigp~at the outskirts of the star-forming clumps first suggests that a fraction of the high gas velocity dispersions in clumpy galaxies may be attributable to some sort of interaction between clumps and the surrounding disc gas. Moreover this enhanced velocity dispersion at the edges of clumps can be explained by several physical mechanisms:  a) turbulence and/or heat radiation feedback which would provide pressure support to the clumps and result in massive outflows \citep{Krumholz_2010}. b) Inflows, which would create a range of velocity vectors at the edge of clumps. c) Internal dynamics of clumps decoupled from the dynamics of the disc.

Outflows can be traced via the H$\alpha$ emission which contains information from ionized gas and shocked stellar winds.  Previous observations of H$\alpha$ kinematics have seen these outflows through a broad component in the H$\alpha$ emission line \citep{Genzel_2011, Wisnioski_2012,Newman_2012a}. However, this broad component is not seen in the \paa~line emission of our observations.

\citet{Newman_2012a} investigated the different sources of high disc velocity dispersion in turbulent galaxies at high redshifts. They found that the gas pressure produced by turbulence and radiation pressure is less than the gas pressure expected from the velocity dispersion measured in these turbulent discs. Suggesting that although star formation feedback is an important source of velocity dispersion, another strong component would be necessary to match the velocity dispersion observed. Our observations match this hypothesis, as the star formation feedback does not seem to fully explain the velocity dispersion maps.   

These results reinforce the question: Do star-forming clumps live long enough to survive several dynamical times of the disc and spiral to the central overdensity? The existence of strong outflows could suggest that the radiation pressure could disrupt the clumps before virialization \citep{Genel_2012}. \citet{Bournaud_2014} stated that the most massive clumps survive to stellar winds due to the gas rich environment where they live, which allows them to recover their mass and live long enough to merge into the galaxy bulge. Using the DYNAMO-OSIRIS sample we will address this question in a following paper, by measuring the velocity sheers of the clumps in the galaxies with the highest spatial resolution (2-3 galaxies, $z<0.15$).

\section{Conclusions}
We presented the high resolution observations of 7 low redshift turbulent discs ($0.07<z<0.2$; DYNAMO-OSIRIS), analogous to high redshift galaxies ($z\sim1.5$). These galaxies were observed with the OSIRIS IFS on Keck. We use the \paa~emission line to construct \sigsfr~and kinematic maps. The galaxies have prominent star-forming clumps (3 to 7 per galaxies) and have disc like kinematics. 
\\\\
To construct reliable velocity dispersion maps we account for several systematics: spatial sampling, AO PSF and noise effects. The spatial sampling correction is done by fitting an analytical function \citep{Epinat_2008a} to the galaxies to later subtract the velocity gradient per spaxel from the observed velocity dispersion. The AO PSF and noise effects  
 are corrected by simulating DYNAMO-OSIRIS-like galaxies and measuring the effect of the AO PSF and noise in different clump configurations. From the simulations, we create systematic maps that we subtract from the observed velocity dispersion.
\\\\
We find that after applying the corrections, there is significant structure in the velocity dispersion maps of turbulent galaxies. To explore this structure, we measure the \sigp~in the velocity dispersion distribution. Around $67$\% of these peaks appear to be connected to the clumps, as they are less than 1~kpc away from the clump centre. We suggest interactions between the clumps and the ISM of the galaxy (outflows, inflows, velocity shears, etc) as possible sources of these \sigp. However, new observations covering a wider area of the galaxy are needed to confirm this result. Further studies will analyze whether these clumps can survive feedback and become virialised.  

\section*{Acknowledgements}

POA acknowledges the tools developed by the Astropy collaboration \citep{Astropy} on the analysis and plotting of IFS data.
Also thanks Javier Zaragoza Cardiel and Sebasti\'an F. Sanch\'ez for the useful discussion on galaxy kinematics. This research was funded by the ARC Discovery Project DP130101460. DBF is the recipient of an Australian Research Council Future Fellowship (FT170100376) funded by the Australian Government. 

%%%%%%%%%%%%%%%%%%%%%%%%%%%%%%%%%%%%%%%%%%%%%%%%%%



\bibliographystyle{mnras}
\bibliography{biblio} % if your bibtex file is called example.bib

\begin{thebibliography}{}
\makeatletter
\relax
\def\mn@urlcharsother{\let\do\@makeother \do\$\do\&\do\#\do\^\do\_\do\%\do\~}
\def\mn@doi{\begingroup\mn@urlcharsother \@ifnextchar [ {\mn@doi@}
  {\mn@doi@[]}}
\def\mn@doi@[#1]#2{\def\@tempa{#1}\ifx\@tempa\@empty \href
  {http://dx.doi.org/#2} {doi:#2}\else \href {http://dx.doi.org/#2} {#1}\fi
  \endgroup}
\def\mn@eprint#1#2{\mn@eprint@#1:#2::\@nil}
\def\mn@eprint@arXiv#1{\href {http://arxiv.org/abs/#1} {{\tt arXiv:#1}}}
\def\mn@eprint@dblp#1{\href {http://dblp.uni-trier.de/rec/bibtex/#1.xml}
  {dblp:#1}}
\def\mn@eprint@#1:#2:#3:#4\@nil{\def\@tempa {#1}\def\@tempb {#2}\def\@tempc
  {#3}\ifx \@tempc \@empty \let \@tempc \@tempb \let \@tempb \@tempa \fi \ifx
  \@tempb \@empty \def\@tempb {arXiv}\fi \@ifundefined
  {mn@eprint@\@tempb}{\@tempb:\@tempc}{\expandafter \expandafter \csname
  mn@eprint@\@tempb\endcsname \expandafter{\@tempc}}}

\bibitem[\protect\citeauthoryear{{Abraham}, {van den Bergh}, {Glazebrook},
  {Ellis}, {Santiago}, {Surma}  \& {Griffiths}}{{Abraham}
  et~al.}{1996}]{Abraham_1996}
{Abraham} R.~G.,  {van den Bergh} S.,  {Glazebrook} K.,  {Ellis} R.~S.,
  {Santiago} B.~X.,  {Surma} P.,   {Griffiths} R.~E.,  1996, \mn@doi [\apjs]
  {10.1086/192352}, \href {http://adsabs.harvard.edu/abs/1996ApJS..107....1A}
  {107, 1}

\bibitem[\protect\citeauthoryear{{Adelman-McCarthy} et~al.,}{{Adelman-McCarthy}
  et~al.}{2006}]{Adelman_2006}
{Adelman-McCarthy} J.~K.,  et~al., 2006, \mn@doi [\apjs] {10.1086/497917},
  \href {http://adsabs.harvard.edu/abs/2006ApJS..162...38A} {162, 38}

\bibitem[\protect\citeauthoryear{{Andersen}, {Bershady}, {Sparke}, {Gallagher},
  {Wilcots}, {van Driel}  \& {Monnier-Ragaigne}}{{Andersen}
  et~al.}{2006}]{Andersen_2006a}
{Andersen} D.~R.,  {Bershady} M.~A.,  {Sparke} L.~S.,  {Gallagher} III J.~S.,
  {Wilcots} E.~M.,  {van Driel} W.,   {Monnier-Ragaigne} D.,  2006, \mn@doi
  [\apjs] {10.1086/506609}, \href
  {http://adsabs.harvard.edu/abs/2006ApJS..166..505A} {166, 505}

\bibitem[\protect\citeauthoryear{{Astropy Collaboration} et~al.,}{{Astropy
  Collaboration} et~al.}{2013}]{Astropy}
{Astropy Collaboration} et~al., 2013, \mn@doi [\aap]
  {10.1051/0004-6361/201322068}, \href
  {http://adsabs.harvard.edu/abs/2013A%26A...558A..33A} {558, A33}

\bibitem[\protect\citeauthoryear{{Barbosa}, {Storchi-Bergmann}, {Cid
  Fernandes}, {Winge}  \& {Schmitt}}{{Barbosa} et~al.}{2006}]{Barbosa_2006}
{Barbosa} F.~K.~B.,  {Storchi-Bergmann} T.,  {Cid Fernandes} R.,  {Winge} C.,
  {Schmitt} H.,  2006, \mn@doi [\mnras] {10.1111/j.1365-2966.2006.10690.x},
  \href {http://adsabs.harvard.edu/abs/2006MNRAS.371..170B} {371, 170}

\bibitem[\protect\citeauthoryear{{Bassett} et~al.,}{{Bassett}
  et~al.}{2014}]{Bassett_2014}
{Bassett} R.,  et~al., 2014, \mn@doi [\mnras] {10.1093/mnras/stu1029}, \href
  {http://adsabs.harvard.edu/abs/2014MNRAS.442.3206B} {442, 3206}

\bibitem[\protect\citeauthoryear{{Bassett} et~al.,}{{Bassett}
  et~al.}{2016}]{Bassett_2016}
{Bassett} R.,  et~al., 2016, preprint, \href
  {http://adsabs.harvard.edu/abs/2016arXiv161105522B} {} (\mn@eprint {arXiv}
  {1611.05522})

\bibitem[\protect\citeauthoryear{{Bekiaris}, {Glazebrook}, {Fluke}  \&
  {Abraham}}{{Bekiaris} et~al.}{2016}]{Bekiaris_2016}
{Bekiaris} G.,  {Glazebrook} K.,  {Fluke} C.~J.,   {Abraham} R.,  2016, \mn@doi
  [\mnras] {10.1093/mnras/stv2292}, \href
  {http://adsabs.harvard.edu/abs/2016MNRAS.455..754B} {455, 754}

\bibitem[\protect\citeauthoryear{{Bournaud} et~al.,}{{Bournaud}
  et~al.}{2008}]{Bournard_2008}
{Bournaud} F.,  et~al., 2008, preprint, \href
  {http://adsabs.harvard.edu/abs/2008arXiv0803.3831B} {803} (\mn@eprint {}
  {0803.3831})

\bibitem[\protect\citeauthoryear{{Bournaud} et~al.,}{{Bournaud}
  et~al.}{2014}]{Bournaud_2014}
{Bournaud} F.,  et~al., 2014, \mn@doi [\apj] {10.1088/0004-637X/780/1/57},
  \href {http://adsabs.harvard.edu/abs/2014ApJ...780...57B} {780, 57}

\bibitem[\protect\citeauthoryear{{Bryant} et~al.,}{{Bryant}
  et~al.}{2015}]{Bryant_2015}
{Bryant} J.~J.,  et~al., 2015, \mn@doi [\mnras] {10.1093/mnras/stu2635}, \href
  {http://adsabs.harvard.edu/abs/2015MNRAS.447.2857B} {447, 2857}

\bibitem[\protect\citeauthoryear{{Calzetti}, {Armus}, {Bohlin}, {Kinney},
  {Koornneef}  \& {Storchi-Bergmann}}{{Calzetti} et~al.}{2000}]{Calzetti_2000}
{Calzetti} D.,  {Armus} L.,  {Bohlin} R.~C.,  {Kinney} A.~L.,  {Koornneef} J.,
   {Storchi-Bergmann} T.,  2000, \mn@doi [\apj] {10.1086/308692}, \href
  {http://adsabs.harvard.edu/abs/2000ApJ...533..682C} {533, 682}

\bibitem[\protect\citeauthoryear{{Ceverino}, {Dekel}  \& {Bournaud}}{{Ceverino}
  et~al.}{2009}]{Ceverino_2009}
{Ceverino} D.,  {Dekel} A.,   {Bournaud} F.,  2009, preprint, \href
  {http://adsabs.harvard.edu/abs/2009arXiv0907.3271C} {} (\mn@eprint {arXiv}
  {0907.3271})

\bibitem[\protect\citeauthoryear{{Ceverino}, {Dekel}, {Mandelker}, {Bournaud},
  {Burkert}, {Genzel}  \& {Primack}}{{Ceverino} et~al.}{2012}]{Ceverino_2012}
{Ceverino} D.,  {Dekel} A.,  {Mandelker} N.,  {Bournaud} F.,  {Burkert} A.,
  {Genzel} R.,   {Primack} J.,  2012, \mn@doi [\mnras]
  {10.1111/j.1365-2966.2011.20296.x}, \href
  {http://adsabs.harvard.edu/abs/2012MNRAS.420.3490C} {420, 3490}

\bibitem[\protect\citeauthoryear{{Coccato}, {Swaters}, {Rubin}, {D'Odorico}  \&
  {McGaugh}}{{Coccato} et~al.}{2008}]{Coccato_2008}
{Coccato} L.,  {Swaters} R.~A.,  {Rubin} V.~C.,  {D'Odorico} S.,   {McGaugh}
  S.~S.,  2008, \mn@doi [\aap] {10.1051/0004-6361:200810410}, \href
  {http://adsabs.harvard.edu/abs/2008A%26A...490..589C} {490, 589}

\bibitem[\protect\citeauthoryear{{Conselice}, {Bershady}  \&
  {Jangren}}{{Conselice} et~al.}{2000}]{Conselice_2000}
{Conselice} C.~J.,  {Bershady} M.~A.,   {Jangren} A.,  2000, \mn@doi [\apj]
  {10.1086/308300}, \href {http://adsabs.harvard.edu/abs/2000ApJ...529..886C}
  {529, 886}

\bibitem[\protect\citeauthoryear{{Courteau}}{{Courteau}}{1997}]{Courteau_1997}
{Courteau} S.,  1997, \mn@doi [\aj] {10.1086/118656}, \href
  {http://adsabs.harvard.edu/abs/1997AJ....114.2402C} {114, 2402}

\bibitem[\protect\citeauthoryear{{Davies}}{{Davies}}{2007}]{Davies_2007}
{Davies} R.~I.,  2007, \mn@doi [\mnras] {10.1111/j.1365-2966.2006.11383.x},
  \href {http://adsabs.harvard.edu/abs/2007MNRAS.375.1099D} {375, 1099}

\bibitem[\protect\citeauthoryear{{Davies} et~al.,}{{Davies}
  et~al.}{2011}]{Davies_2011}
{Davies} R.,  et~al., 2011, \mn@doi [\apj] {10.1088/0004-637X/741/2/69}, \href
  {http://adsabs.harvard.edu/abs/2011ApJ...741...69D} {741, 69}

\bibitem[\protect\citeauthoryear{{Dekel}, {Sari}  \& {Ceverino}}{{Dekel}
  et~al.}{2009}]{Dekel_2009}
{Dekel} A.,  {Sari} R.,   {Ceverino} D.,  2009, \mn@doi [\apj]
  {10.1088/0004-637X/703/1/785}, \href
  {http://adsabs.harvard.edu/abs/2009ApJ...703..785D} {703, 785}

\bibitem[\protect\citeauthoryear{{Dimeo}}{{Dimeo}}{2005}]{Dimeo_2005}
{Dimeo} R.,  2005, PAN User Guide, {ftp://ftp.ncnr.nist.gov/pub/staff/dimeo/
  pandoc.pdf/}

\bibitem[\protect\citeauthoryear{{Dopita}, {Hart}, {McGregor}, {Oates},
  {Bloxham}  \& {Jones}}{{Dopita} et~al.}{2007}]{Dopita_2007}
{Dopita} M.,  {Hart} J.,  {McGregor} P.,  {Oates} P.,  {Bloxham} G.,   {Jones}
  D.,  2007, \mn@doi [\apss] {10.1007/s10509-007-9510-z}, \href
  {http://adsabs.harvard.edu/abs/2007Ap%26SS.310..255D} {310, 255}

\bibitem[\protect\citeauthoryear{{Elmegreen} \& {Elmegreen}}{{Elmegreen} \&
  {Elmegreen}}{2005}]{Elmegreen_2005}
{Elmegreen} B.~G.,  {Elmegreen} D.~M.,  2005, \mn@doi [\apj] {10.1086/430514},
  \href {http://adsabs.harvard.edu/abs/2005ApJ...627..632E} {627, 632}

\bibitem[\protect\citeauthoryear{{Elmegreen}, {Bournaud}  \&
  {Elmegreen}}{{Elmegreen} et~al.}{2008}]{Elmegreen_2008}
{Elmegreen} B.~G.,  {Bournaud} F.,   {Elmegreen} D.~M.,  2008, \mn@doi [\apj]
  {10.1086/592190}, \href {http://adsabs.harvard.edu/abs/2008ApJ...688...67E}
  {688, 67}

\bibitem[\protect\citeauthoryear{{Epinat} et~al.,}{{Epinat}
  et~al.}{2008}]{Epinat_2008a}
{Epinat} B.,  et~al., 2008, \mn@doi [\mnras]
  {10.1111/j.1365-2966.2008.13422.x}, \href
  {http://adsabs.harvard.edu/abs/2008MNRAS.388..500E} {388, 500}

\bibitem[\protect\citeauthoryear{{Epinat}, {Amram}, {Balkowski}  \&
  {Marcelin}}{{Epinat} et~al.}{2010}]{Epinat_2010}
{Epinat} B.,  {Amram} P.,  {Balkowski} C.,   {Marcelin} M.,  2010, \mn@doi
  [\mnras] {10.1111/j.1365-2966.2009.15688.x}, \href
  {http://adsabs.harvard.edu/abs/2010MNRAS.401.2113E} {401, 2113}

\bibitem[\protect\citeauthoryear{{Fisher} et~al.,}{{Fisher}
  et~al.}{2014}]{Fisher_2014}
{Fisher} D.~B.,  et~al., 2014, \mn@doi [\apjl] {10.1088/2041-8205/790/2/L30},
  \href {http://adsabs.harvard.edu/abs/2014ApJ...790L..30F} {790, L30}

\bibitem[\protect\citeauthoryear{{Fisher} et~al.,}{{Fisher}
  et~al.}{2017}]{Fisher_2016}
{Fisher} D.~B.,  et~al., 2017, \mn@doi [\mnras] {10.1093/mnras/stw2281}, \href
  {http://adsabs.harvard.edu/abs/2017MNRAS.464..491F} {464, 491}

\bibitem[\protect\citeauthoryear{{F{\"o}rster Schreiber} et~al.,}{{F{\"o}rster
  Schreiber} et~al.}{2006}]{Forster_2006}
{F{\"o}rster Schreiber} N.~M.,  et~al., 2006, \mn@doi [\apj] {10.1086/504403},
  \href {http://adsabs.harvard.edu/abs/2006ApJ...645.1062F} {645, 1062}

\bibitem[\protect\citeauthoryear{{F{\"o}rster Schreiber} et~al.,}{{F{\"o}rster
  Schreiber} et~al.}{2009}]{Forster_2009}
{F{\"o}rster Schreiber} N.~M.,  et~al., 2009, \mn@doi [\apj]
  {10.1088/0004-637X/706/2/1364}, \href
  {http://adsabs.harvard.edu/abs/2009ApJ...706.1364F} {706, 1364}

\bibitem[\protect\citeauthoryear{{Genel} et~al.,}{{Genel}
  et~al.}{2012}]{Genel_2012}
{Genel} S.,  et~al., 2012, \mn@doi [\apj] {10.1088/0004-637X/745/1/11}, \href
  {http://adsabs.harvard.edu/abs/2012ApJ...745...11G} {745, 11}

\bibitem[\protect\citeauthoryear{{Genzel} et~al.,}{{Genzel}
  et~al.}{2006}]{Genzel_2006}
{Genzel} R.,  et~al., 2006, \mn@doi [\nat] {10.1038/nature05052}, \href
  {http://adsabs.harvard.edu/abs/2006Natur.442..786G} {442, 786}

\bibitem[\protect\citeauthoryear{{Genzel} et~al.,}{{Genzel}
  et~al.}{2011}]{Genzel_2011}
{Genzel} R.,  et~al., 2011, \mn@doi [\apj] {10.1088/0004-637X/733/2/101}, \href
  {http://adsabs.harvard.edu/abs/2011ApJ...733..101G} {733, 101}

\bibitem[\protect\citeauthoryear{{Glazebrook}}{{Glazebrook}}{2013}]{Glazebrook_2013}
{Glazebrook} K.,  2013, \mn@doi [\pasa] {10.1017/pasa.2013.34}, \href
  {http://adsabs.harvard.edu/abs/2013PASA...30...56G} {30, e056}

\bibitem[\protect\citeauthoryear{{Green} et~al.,}{{Green}
  et~al.}{2010}]{Green_2010}
{Green} A.~W.,  et~al., 2010, \mn@doi [\nat] {10.1038/nature09452}, \href
  {http://adsabs.harvard.edu/abs/2010Natur.467..684G} {467, 684}

\bibitem[\protect\citeauthoryear{{Green} et~al.,}{{Green}
  et~al.}{2014}]{Green_2014}
{Green} A.~W.,  et~al., 2014, \mn@doi [\mnras] {10.1093/mnras/stt1882}, \href
  {http://adsabs.harvard.edu/abs/2014MNRAS.437.1070G} {437, 1070}

\bibitem[\protect\citeauthoryear{{Gruyters}, {Exter}, {Roberts}  \&
  {Rappaport}}{{Gruyters} et~al.}{2012}]{Gruyters_2012}
{Gruyters} P.,  {Exter} K.,  {Roberts} T.~P.,   {Rappaport} S.,  2012, \mn@doi
  [\aap] {10.1051/0004-6361/201219051}, \href
  {http://adsabs.harvard.edu/abs/2012A%26A...544A..86G} {544, A86}

\bibitem[\protect\citeauthoryear{{Guo} et~al.,}{{Guo} et~al.}{2015}]{Guo_2015}
{Guo} Y.,  et~al., 2015, \mn@doi [\apj] {10.1088/0004-637X/800/1/39}, \href
  {http://adsabs.harvard.edu/abs/2015ApJ...800...39G} {800, 39}

\bibitem[\protect\citeauthoryear{{Hao}, {Kennicutt}, {Johnson}, {Calzetti},
  {Dale}  \& {Moustakas}}{{Hao} et~al.}{2011}]{Hao_2011}
{Hao} C.-N.,  {Kennicutt} R.~C.,  {Johnson} B.~D.,  {Calzetti} D.,  {Dale}
  D.~A.,   {Moustakas} J.,  2011, \mn@doi [\apj] {10.1088/0004-637X/741/2/124},
  \href {http://adsabs.harvard.edu/abs/2011ApJ...741..124H} {741, 124}

\bibitem[\protect\citeauthoryear{{Immeli}, {Samland}, {Gerhard}  \&
  {Westera}}{{Immeli} et~al.}{2004a}]{Immeli_2004}
{Immeli} A.,  {Samland} M.,  {Gerhard} O.,   {Westera} P.,  2004a, \mn@doi
  [\aap] {10.1051/0004-6361:20034282}, \href
  {http://adsabs.harvard.edu/abs/2004A%26A...413..547I} {413, 547}

\bibitem[\protect\citeauthoryear{{Immeli}, {Samland}, {Westera}  \&
  {Gerhard}}{{Immeli} et~al.}{2004b}]{Immeli_2004b}
{Immeli} A.,  {Samland} M.,  {Westera} P.,   {Gerhard} O.,  2004b, \mn@doi
  [\apj] {10.1086/422179}, \href
  {http://adsabs.harvard.edu/abs/2004ApJ...611...20I} {611, 20}

\bibitem[\protect\citeauthoryear{{Jones}, {Swinbank}, {Ellis}, {Richard}  \&
  {Stark}}{{Jones} et~al.}{2010}]{Jones_2010}
{Jones} T.~A.,  {Swinbank} A.~M.,  {Ellis} R.~S.,  {Richard} J.,   {Stark}
  D.~P.,  2010, \mn@doi [\mnras] {10.1111/j.1365-2966.2010.16378.x}, \href
  {http://adsabs.harvard.edu/abs/2010MNRAS.404.1247J} {404, 1247}

\bibitem[\protect\citeauthoryear{{Kauffmann} et~al.,}{{Kauffmann}
  et~al.}{2003}]{Kauffmann_2003}
{Kauffmann} G.,  et~al., 2003, \mn@doi [\mnras]
  {10.1046/j.1365-8711.2003.06291.x}, \href
  {http://adsabs.harvard.edu/abs/2003MNRAS.341...33K} {341, 33}

\bibitem[\protect\citeauthoryear{{Krumholz} \& {Dekel}}{{Krumholz} \&
  {Dekel}}{2010}]{Krumholz_2010}
{Krumholz} M.~R.,  {Dekel} A.,  2010, \mn@doi [\mnras]
  {10.1111/j.1365-2966.2010.16675.x}, \href
  {http://adsabs.harvard.edu/abs/2010MNRAS.406..112K} {406, 112}

\bibitem[\protect\citeauthoryear{{Larkin} et~al.,}{{Larkin}
  et~al.}{2006}]{Larkin_2006}
{Larkin} J.,  et~al., 2006, \mn@doi [\nar] {10.1016/j.newar.2006.02.005}, \href
  {http://adsabs.harvard.edu/abs/2006NewAR..50..362L} {50, 362}

\bibitem[\protect\citeauthoryear{{Law}, {Steidel}  \& {Erb}}{{Law}
  et~al.}{2006}]{Law_2006}
{Law} D.~R.,  {Steidel} C.~C.,   {Erb} D.~K.,  2006, \mn@doi [\aj]
  {10.1086/498683}, \href {http://adsabs.harvard.edu/abs/2006AJ....131...70L}
  {131, 70}

\bibitem[\protect\citeauthoryear{{Law}, {Steidel}, {Erb}, {Larkin}, {Pettini},
  {Shapley}  \& {Wright}}{{Law} et~al.}{2009}]{Law_2009}
{Law} D.~R.,  {Steidel} C.~C.,  {Erb} D.~K.,  {Larkin} J.~E.,  {Pettini} M.,
  {Shapley} A.~E.,   {Wright} S.~A.,  2009, \mn@doi [\apj]
  {10.1088/0004-637X/697/2/2057}, \href
  {http://adsabs.harvard.edu/abs/2009ApJ...697.2057L} {697, 2057}

\bibitem[\protect\citeauthoryear{{Livermore} et~al.,}{{Livermore}
  et~al.}{2012}]{Livermore_2012}
{Livermore} R.~C.,  et~al., 2012, \mn@doi [\mnras]
  {10.1111/j.1365-2966.2012.21900.x}, \href
  {http://adsabs.harvard.edu/abs/2012MNRAS.427..688L} {427, 688}

\bibitem[\protect\citeauthoryear{{Livermore} et~al.,}{{Livermore}
  et~al.}{2015}]{Livermore_2015}
{Livermore} R.~C.,  et~al., 2015, \mn@doi [\mnras] {10.1093/mnras/stv686},
  \href {http://adsabs.harvard.edu/abs/2015MNRAS.450.1812L} {450, 1812}

\bibitem[\protect\citeauthoryear{{Mieda}, {Wright}, {Larkin}, {Armus},
  {Juneau}, {Salim}  \& {Murray}}{{Mieda} et~al.}{2016}]{Mieda_2016}
{Mieda} E.,  {Wright} S.~A.,  {Larkin} J.~E.,  {Armus} L.,  {Juneau} S.,
  {Salim} S.,   {Murray} N.,  2016, \mn@doi [\apj]
  {10.3847/0004-637X/831/1/78}, \href
  {http://adsabs.harvard.edu/abs/2016ApJ...831...78M} {831, 78}

\bibitem[\protect\citeauthoryear{{Newman} et~al.,}{{Newman}
  et~al.}{2012a}]{Newman_2012a}
{Newman} S.~F.,  et~al., 2012a, \mn@doi [\apj] {10.1088/0004-637X/752/2/111},
  \href {http://adsabs.harvard.edu/abs/2012ApJ...752..111N} {752, 111}

\bibitem[\protect\citeauthoryear{{Newman} et~al.,}{{Newman}
  et~al.}{2012b}]{Newman_2012b}
{Newman} S.~F.,  et~al., 2012b, \mn@doi [\apj] {10.1088/0004-637X/761/1/43},
  \href {http://adsabs.harvard.edu/abs/2012ApJ...761...43N} {761, 43}

\bibitem[\protect\citeauthoryear{{Obreschkow} et~al.,}{{Obreschkow}
  et~al.}{2015}]{Obreschkow_2015}
{Obreschkow} D.,  et~al., 2015, preprint, \href
  {http://adsabs.harvard.edu/abs/2015arXiv150804768O} {} (\mn@eprint {arXiv}
  {1508.04768})

\bibitem[\protect\citeauthoryear{{Rousselot}, {Lidman}, {Cuby}, {Moreels}  \&
  {Monnet}}{{Rousselot} et~al.}{2000}]{Rousselot_2000}
{Rousselot} P.,  {Lidman} C.,  {Cuby} J.-G.,  {Moreels} G.,   {Monnet} G.,
  2000, \aap, \href {http://adsabs.harvard.edu/abs/2000A%26A...354.1134R} {354,
  1134}

\bibitem[\protect\citeauthoryear{{Shapiro} et~al.,}{{Shapiro}
  et~al.}{2008}]{Shapiro_2008}
{Shapiro} K.~L.,  et~al., 2008, \mn@doi [\apj] {10.1086/587133}, \href
  {http://adsabs.harvard.edu/abs/2008ApJ...682..231S} {682, 231}

\bibitem[\protect\citeauthoryear{{Sharp} et~al.,}{{Sharp}
  et~al.}{2006}]{Sharp_2006}
{Sharp} R.,  et~al., 2006, in Society of Photo-Optical Instrumentation
  Engineers (SPIE) Conference Series. p. 62690G (\mn@eprint {}
  {astro-ph/0606137}), \mn@doi{10.1117/12.671022}

\bibitem[\protect\citeauthoryear{{Skilling}}{{Skilling}}{2004}]{Skilling_2004}
{Skilling} J.,  2004, in {Fischer} R.,  {Preuss} R.,   {Toussaint} U.~V.,  eds,
   American Institute of Physics Conference Series Vol. 735, American Institute
  of Physics Conference Series. pp 395--405, \mn@doi{10.1063/1.1835238}

\bibitem[\protect\citeauthoryear{{Skrutskie} et~al.,}{{Skrutskie}
  et~al.}{2006}]{Skrutskie_2006}
{Skrutskie} M.~F.,  et~al., 2006, \mn@doi [\aj] {10.1086/498708}, \href
  {http://adsabs.harvard.edu/abs/2006AJ....131.1163S} {131, 1163}

\bibitem[\protect\citeauthoryear{{Swinbank}, {Smail}, {Sobral}, {Theuns},
  {Best}  \& {Geach}}{{Swinbank} et~al.}{2012}]{Swinbank_2012}
{Swinbank} A.~M.,  {Smail} I.,  {Sobral} D.,  {Theuns} T.,  {Best} P.~N.,
  {Geach} J.~E.,  2012, \mn@doi [\apj] {10.1088/0004-637X/760/2/130}, \href
  {http://adsabs.harvard.edu/abs/2012ApJ...760..130S} {760, 130}

\bibitem[\protect\citeauthoryear{{Tamburello}, {Rahmati}, {Mayer}, {Cava},
  {Dessauges-Zavadsky}  \& {Schaerer}}{{Tamburello}
  et~al.}{2016}]{Tamburello_2016}
{Tamburello} V.,  {Rahmati} A.,  {Mayer} L.,  {Cava} A.,  {Dessauges-Zavadsky}
  M.,   {Schaerer} D.,  2016, preprint, \href
  {http://adsabs.harvard.edu/abs/2016arXiv161005304T} {} (\mn@eprint {arXiv}
  {1610.05304})

\bibitem[\protect\citeauthoryear{{Varidel}, {Pracy}, {Croom}, {Owers}  \&
  {Sadler}}{{Varidel} et~al.}{2016}]{Varidel_2016}
{Varidel} M.,  {Pracy} M.,  {Croom} S.,  {Owers} M.~S.,   {Sadler} E.,  2016,
  \mn@doi [\pasa] {10.1017/pasa.2016.3}, \href
  {http://adsabs.harvard.edu/abs/2016PASA...33....6V} {33, e006}

\bibitem[\protect\citeauthoryear{{Westmoquette}, {Exter}, {Smith}  \&
  {Gallagher}}{{Westmoquette} et~al.}{2007}]{Westmoquette_2007}
{Westmoquette} M.~S.,  {Exter} K.~M.,  {Smith} L.~J.,   {Gallagher} J.~S.,
  2007, \mn@doi [\mnras] {10.1111/j.1365-2966.2007.12346.x}, \href
  {http://adsabs.harvard.edu/abs/2007MNRAS.381..894W} {381, 894}

\bibitem[\protect\citeauthoryear{{Westmoquette}, {Smith}  \&
  {Gallagher}}{{Westmoquette} et~al.}{2008}]{Westmoquette_2008}
{Westmoquette} M.~S.,  {Smith} L.~J.,   {Gallagher} J.~S.,  2008, \mn@doi
  [\mnras] {10.1111/j.1365-2966.2007.12628.x}, \href
  {http://adsabs.harvard.edu/abs/2008MNRAS.383..864W} {383, 864}

\bibitem[\protect\citeauthoryear{{Westmoquette}, {Clements}, {Bendo}  \&
  {Khan}}{{Westmoquette} et~al.}{2012}]{Westmoquette_2012}
{Westmoquette} M.~S.,  {Clements} D.~L.,  {Bendo} G.~J.,   {Khan} S.~A.,  2012,
  \mn@doi [\mnras] {10.1111/j.1365-2966.2012.21214.x}, \href
  {http://adsabs.harvard.edu/abs/2012MNRAS.424..416W} {424, 416}

\bibitem[\protect\citeauthoryear{{Wisnioski} et~al.,}{{Wisnioski}
  et~al.}{2011}]{Wisnioski_2011}
{Wisnioski} E.,  et~al., 2011, \mn@doi [\mnras]
  {10.1111/j.1365-2966.2011.19429.x}, \href
  {http://adsabs.harvard.edu/abs/2011MNRAS.417.2601W} {417, 2601}

\bibitem[\protect\citeauthoryear{{Wisnioski}, {Glazebrook}, {Blake}, {Poole},
  {Green}, {Wyder}  \& {Martin}}{{Wisnioski} et~al.}{2012}]{Wisnioski_2012}
{Wisnioski} E.,  {Glazebrook} K.,  {Blake} C.,  {Poole} G.~B.,  {Green} A.~W.,
  {Wyder} T.,   {Martin} C.,  2012, \mn@doi [\mnras]
  {10.1111/j.1365-2966.2012.20850.x}, \href
  {http://adsabs.harvard.edu/abs/2012MNRAS.422.3339W} {422, 3339}

\bibitem[\protect\citeauthoryear{{Wizinowich} et~al.,}{{Wizinowich}
  et~al.}{2006}]{Wizinowich_2006}
{Wizinowich} P.~L.,  et~al., 2006, in Society of Photo-Optical Instrumentation
  Engineers (SPIE) Conference Series. p. 627209, \mn@doi{10.1117/12.672104}

\bibitem[\protect\citeauthoryear{{Wuyts} et~al.,}{{Wuyts}
  et~al.}{2012}]{Wuyts_2012}
{Wuyts} S.,  et~al., 2012, \mn@doi [\apj] {10.1088/0004-637X/753/2/114}, \href
  {http://adsabs.harvard.edu/abs/2012ApJ...753..114W} {753, 114}

\makeatother
\end{thebibliography}

%\begin{thebibliography}{99}
%\end{thebibliography}

%%%%%%%%%%%%%%%%%%%%%%%%%%%%%%%%%%%%%%%%%%%%%%%%%%

%%%%%%%%%%%%%%%%% APPENDICES %%%%%%%%%%%%%%%%%%%%%

%%%%%%%%%%%%%%%%%%%%%%%%%%%%%%%%%%%%%%%%%%%%%%%%%%

% Don't change these lines
\bsp	% typesetting comment
\label{lastpage}
\end{document}